\documentclass[pra,longbibliography,twocolumn,showpacs,nofootinbib,superscriptaddress,notitlepage]{revtex4-1}

\pdfoutput=1 

\usepackage{dsfont}
\usepackage{amsmath}
\usepackage{amssymb,bm}
\usepackage{amsthm}
\usepackage{mathtools}
\usepackage{url}

\usepackage{array, makecell}
\usepackage{boldline}
\usepackage{enumitem}

\usepackage{color,dsfont}
\usepackage{graphicx}
\usepackage{ragged2e}
\usepackage[colorlinks=true, hyperindex, breaklinks, linkcolor=blue, urlcolor=blue, citecolor=blue]{hyperref} 
\usepackage[normalem]{ulem}
\usepackage[capitalise]{cleveref}
\usepackage{mathrsfs}
\usepackage[caption=false]{subfig}
\usepackage{mathtools}
\usepackage{float}
\usepackage{verbatim}
\usepackage{latexsym}
\usepackage{amsmath}
\usepackage{amssymb}
\usepackage{setspace}
\usepackage{amsfonts}
\usepackage{stmaryrd}
\usepackage{xcolor}
\usepackage{enumitem}
\usepackage{lipsum}

\begin{document}
\title{Fault-tolerant bosonic quantum error correction with the surface-GKP code} 
\author{Kyungjoo Noh}\email{noh827@gmail.com}
\affiliation{Department of Physics, Yale University, New Haven, CT, 06520, United States}
\author{Christopher Chamberland}\email{mathematicschris@gmail.com}
\affiliation{IBM T. J. Watson Research Center, Yorktown Heights, NY, 10598, United States}
\begin{abstract}
Bosonic quantum error correction is a viable option for realizing error-corrected quantum information processing in continuous-variable bosonic systems. Various single-mode bosonic quantum error-correcting codes such as cat, binomial, and GKP codes have been implemented experimentally in circuit QED and trapped ion systems. Moreover, there have been many theoretical proposals to scale up such single-mode bosonic codes to realize large-scale fault-tolerant quantum computation. Here, we consider the concatenation of the single-mode GKP code with the surface code, namely, the surface-GKP code. In particular, we thoroughly investigate the performance of the surface-GKP code by assuming realistic GKP states with a finite squeezing and noisy circuit elements due to photon losses. By using a minimum-weight perfect matching decoding algorithm on a 3D space-time graph, we show that fault-tolerant quantum error correction is possible with the surface-GKP code if the squeezing of the GKP states is higher than $11.2$dB in the case where the GKP states are the only noisy elements. We also show that the squeezing threshold changes to $18.6$dB when both the GKP states and circuit elements are comparably noisy. At this threshold, each circuit component fails with probability $0.69\%$. Finally, if the GKP states are noiseless, fault-tolerant quantum error correction with the surface-GKP code is possible if each circuit element fails with probability less than $0.81\%$. We stress that our decoding scheme uses the additional information from GKP-stabilizer measurements and we provide a simple method to compute renormalized edge weights of the matching graphs. Furthermore, our noise model is general as it includes full circuit-level noise.  
\end{abstract}
\maketitle

\section{Introduction}
\label{section:Introduction}

Continuous-variable systems or bosonic modes are ubiquitous in many quantum computing platforms and there have been various proposals for realizing quantum computation in continuous-variable systems \cite{Lloyd1999,Jeong2002,Ralph2003,Lund2008}. Notably, bosonic quantum error correction \cite{Albert2018} has recently risen as a hardware-efficient route to implement quantum error correction (QEC) by taking advantage of the infinite-dimensionality of a bosonic Hilbert space. Various bosonic quantum error-correcting codes include Schr\"odinger's cat \cite{Cochrane1999}, binomial \cite{Michael2016}, and Gottesman-Kitaev-Preskill (GKP) \cite{Gottesman2001} codes. All these codes encode a logical qubit in a physical bosonic oscillator mode and have been realized experimentally in circuit QED \cite{Leghtas2015,Ofek2016,Touzard2018,Hu2019,Campagne2019,Grimm2019} and trapped ion \cite{Fluhmann2018,Fluhmann2019,Fluhmann2019b} systems in the past few years.

While bosonic QEC with a single bosonic mode (and a single ancilla qubit) can suppress relevant errors such as photon losses or phase space shift errors in a hardware-efficient way, it should also be noted that logical error rates cannot be suppressed to an arbitrarily small value with this minimal architecture. For example, the experimentally realized four-component cat code and the binomial code cannot correct two (or more) photon loss events. Similarly, the GKP code cannot correct phase space shift errors of a size larger than a critical value. Therefore, to further suppress the residual errors, these bosonic codes should, for example, be concatenated with some other error-correcting code families such as the surface code \cite{Bravyi1998,Dennis2002,Fowler2012}.

Recently, there have been proposals for scaling up the cat codes by concatenating them with a repetition code \cite{Guillaud2019} or a surface code \cite{Puri2019} which are tailored to biased noise models \cite{Tuckett2018,Tuckett2018b,Tuckett2019}. These schemes take advantage of the fact that the cat code can suppress bosonic dephasing (stochastic random rotation) errors exponentially in the size of the cat code, thereby yielding a qubit with a biased noise predominated either by bit-flip or phase-flip errors. These studies have shown that the gates on the cat code needed for the concatenation can be implemented in a noise-bias-preserving way. On the other hand, the full concatenated error correction schemes have not been thoroughly studied in these works.

Meanwhile, there have also been studies on scaling up the GKP code by concatenating it with a repetition code \cite{Fukui2017}, the $[[4,2,2]]$ code \cite{Fukui2017,Fukui2018b}, and the surface code \cite{Wang2017,Fukui2018a,Vuillot2019}, or by using cluster states and measurement-based quantum computation \cite{Menicucci2014,Fukui2018a,Fukui2019}. One of the recurring themes in these previous works is that the continuous error information gathered during the GKP code error correction protocol can boost the performance of the next layer of the concatenated error correction. For example, while the surface code by itself has the code capacity threshold $\sim \!\! 11\%$ \cite{Dennis2002}, the threshold can be increased to $\sim \!\! 14\%$ if the additional error information from GKP-stabilizer measurements is incorporated in the surface code error correction protocol \cite{Wang2017,Fukui2018a,Vuillot2019}. However, note that the code capacity thresholds are obtained by assuming that only qubits can fail, i.e., gates, state preparations, and measurements are assumed perfect. Hence, the above code capacity threshold for the concatenated GKP code is evaluated by assuming noiseless GKP and surface code stabilizer measurements, or equivalently, by assuming that ideal GKP states (with an infinitely large squeezing) are used for the stabilizer measurements.

If the error syndrome is extracted using realistic GKP states with a finite squeezing, the error correction protocols would become faulty. Nevertheless, in the framework of measurement-based quantum computation \cite{Raussendorf2003}, it has been shown that fault-tolerant quantum error correction with finitely-squeezed GKP states is possible if the strength of the squeezing is above a certain threshold. Specifically, the recent works \cite{Fukui2018a,Fukui2019} have demonstrated that the threshold value can be brought down from $\sim\!\! 20$dB \cite{Menicucci2014} to less than $10$dB by using post-selection.

In the framework of gate-based quantum computation, several fault-tolerance thresholds have been computed for the GKP code concatenated with the toric code, namely the toric-GKP code, by assuming a phenomenological noise model \cite{Vuillot2019,Wang2017}. In these previous works, however, shift errors were manually added instead of being derived from an underlying noise model for realistic GKP states and the noisy circuits used for stabilizer measurements.

In our work, we thoroughly investigate the full error correction protocol for the GKP code concatenated with the surface code, namely, the surface-GKP code. We choose the surface code for the next level of encoding because it can be implemented in a geometrically local way in a planar architecture. In particular, we consider a detailed circuit-level noise model and assume that every GKP state supplied to the error correction chain is finitely squeezed, and also that every circuit element can be noisy due to photon losses and heating. Unlike previous works such as in \cite{Vuillot2019,Wang2017} (where noise propagation was not considered), we comprehensively take into account the propagation of such imperfections throughout the entire circuit and simulate the full surface code error correction protocol assuming this general circuit-level noise model. 
Finally, by using a simple decoding algorithm based on a minimum-weight perfect matching (MWPM) \cite{Edmonds1965,Edmonds1965b} 
algorithm applied to 3D space-time graphs, we establish that fault-tolerant quantum error correction is possible if the squeezing of the GKP states is higher than $11.2$dB when the GKP states are the only noisy components, or than $18.6$dB when both the GKP states and circuit elements are comparably noisy. In the latter case, each circuit element that implements the surface-GKP code fails with probability $0.69\%$. In the case where GKP states are noiseless, we find that fault-tolerant quantum error correction with the surface-GKP code is possible if each circuit element fails with probability less than $0.81\%$. In general, it has been shown that using edge weights in the matching graphs which are computed from the most likely error configurations can significantly improve the performance of a topological code \cite{Wang2011,ChamberlandFlagToric2019}. Our decoding algorithm provides a simple way to compute renormalized edge weights of the 3D matching graphs, tailored to our general circuit-level noise model, based on information obtained from GKP-stabilizer measurements. 

Our paper is organized as follows: In \cref{section:Surface-GKP codes}, we introduce the surface-GKP code and describe the noise model that we assume for the fault-tolerance study. In \cref{section:Main results}, we summarize the main results and establish fault-tolerance thresholds. A detailed description of our analysis is given in \cref{section:Methods}. In \cref{section:Discussion}, we compare our results with the previous ones and conclude the paper with an outlook.

\section{The surface-GKP code} 
\label{section:Surface-GKP codes}

In this section, we introduce the surface-GKP code, i.e., GKP qubits concatenated with the surface code. The GKP qubits are constructed by using the standard square-lattice GKP code that encodes a single qubit into an oscillator mode \cite{Gottesman2001}, which is reviewed in \cref{subsection:GKP qubit}. For the next layer of the encoding, we use the family of rotated surface codes that requires $d^{2}$ data qubits and $d^{2}-1$ syndrome qubits where $d \in \lbrace 2n+1 : n\in \mathbb{N}  \rbrace$ is the distance of the code \cite{Bombin2007,Tomita2014}. In \cref{subsection:Surface codes with GKP qubits}, we construct the surface-GKP code and discuss its implementation. In \cref{subsection:Noise models}, we introduce the noise model that we use to simulate the full noisy error correction protocol for the surface-GKP code. Readers who are familiar with the GKP code and the surface code may skip \cref{subsection:GKP qubit,subsection:Surface codes with GKP qubits} and are referred to \cref{subsection:Noise models}. 

\subsection{GKP qubit}
\label{subsection:GKP qubit}

\begin{figure*}[t!]
\centering
\includegraphics[width=6.7in]{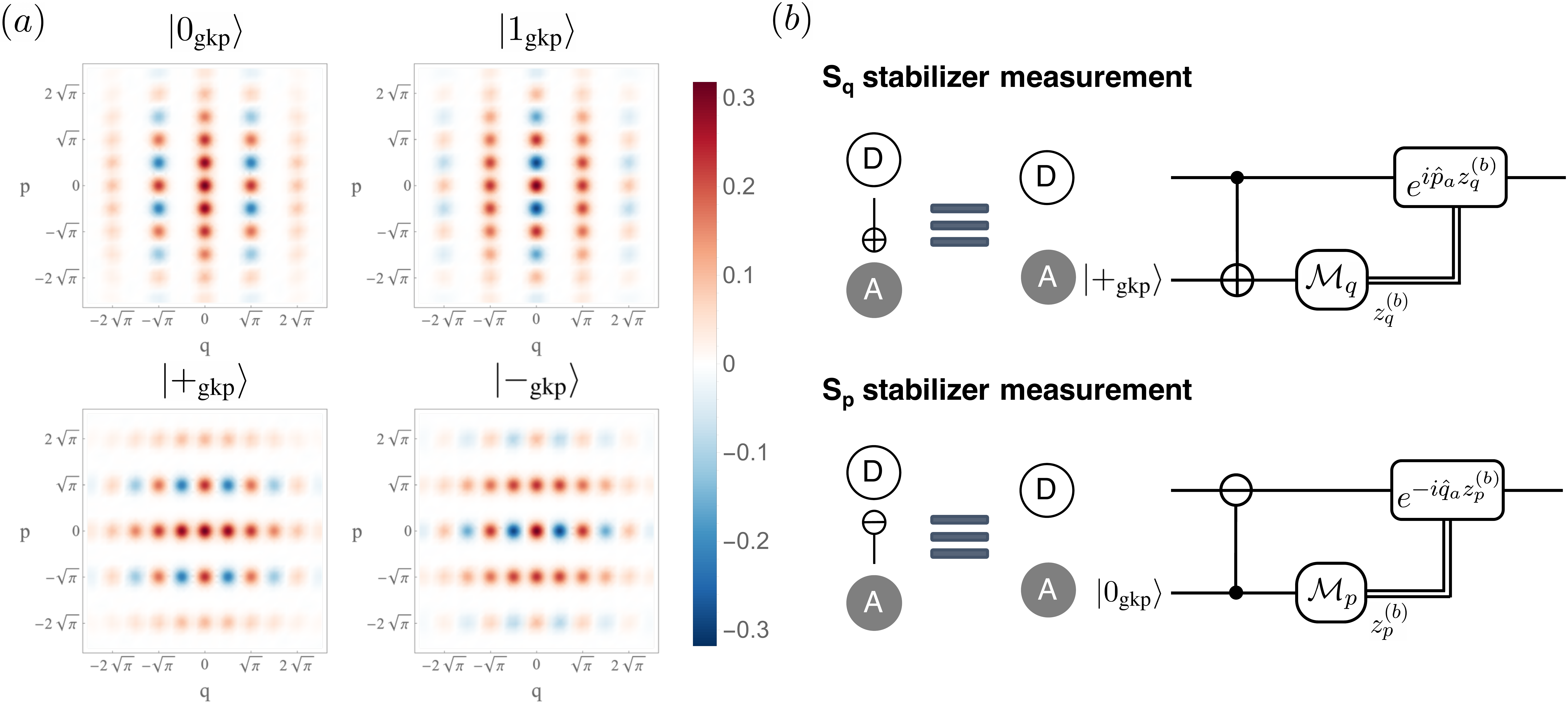}
\caption{(a) Computational basis states ($|0_{\textrm{gkp}}\rangle$, $|1_{\textrm{gkp}}\rangle$) and complementary basis states ($|+_{\textrm{gkp}}\rangle$, $|-_{\textrm{gkp}}\rangle$) of an approximate GKP qubit with an average photon number $\bar{n}=5$. (b) Circuits for measuring the $\hat{S}_{q}$ and $\hat{S}_{p}$ stabilizers. $\mathcal{M}_{q}$ and $\mathcal{M}_{p}$ represent the homodyne measurement of the position and momentum operators, respectively. Also, the controlled-$\oplus$ symbol represents the SUM gate and similarly the controlled-$\ominus$ symbol represents the inverse-SUM gate (see \cref{eq:Clifford gates for GKP qubits}). Note that the size of the correction shifts $\exp[i\hat{p}_{a}z_{q}^{(b)}]$ and $\exp[-i\hat{q}_{a}z_{p}^{(b)}]$ in the $\hat{S}_{q}$ and $\hat{S}_{p}$ stabilizer measurements are determined by the homodyne measurement outcomes $z_{q}^{(b)}$ and $z_{p}^{(b)}$.    }
\label{fig:GKP qubit fundamentals}
\end{figure*}

Let $\hat{q} = (\hat{a}^{\dagger}+\hat{a})/\sqrt{2}$ and $\hat{p} = i(\hat{a}^{\dagger}-\hat{a})/\sqrt{2}$ be the position and momentum operators of a bosonic mode, where $\hat{a}$ and $\hat{a}^{\dagger}$ are annihilation and creation operators satisfying $[\hat{a},\hat{a}^{\dagger}]=1$. We define the GKP qubit as the $2$-dimensional subspace of a bosonic Hilbert space that is stabilized by the two stabilizers 
\begin{align}
\hat{S}_{q} \equiv \exp[ i2\sqrt{\pi}\hat{q} ] , \quad \hat{S}_{p} \equiv \exp[ -i2\sqrt{\pi}\hat{p} ] . 
\end{align}
Measuring these two commuting stabilizers is equivalent to measuring the position and momentum operators $\hat{q}$ and $\hat{p}$ modulo $\sqrt{\pi}$. Therefore, any phase space shift error $\exp[i(\xi_{p}\hat{q} - \xi_{q}\hat{p})]$ acting on the ideal GKP qubit can be detected and corrected as long as $|\xi_{q}|,|\xi_{p}| < \sqrt{\pi}/2$.

Explicitly, the computational basis states of the ideal GKP qubit are given by 
\begin{align}
|0_{\textrm{gkp}}\rangle &= \sum_{n\in\mathbb{Z}} |\hat{q} = 2n\sqrt{\pi} \rangle,
\nonumber\\
|1_{\textrm{gkp}}\rangle &= \sum_{n\in\mathbb{Z}} |\hat{q} = (2n+1)\sqrt{\pi} \rangle. 
\end{align}
Also, the complementary basis states $|\pm_{\textrm{gkp}}\rangle \equiv \frac{1}{\sqrt{2}}(|0_{\textrm{gkp}}\rangle \pm|1_{\textrm{gkp}}\rangle)$ are given by 
\begin{align}
|+_{\textrm{gkp}}\rangle &=  \sum_{n\in\mathbb{Z}} |\hat{p} = 2n\sqrt{\pi} \rangle,
\nonumber\\
|-_{\textrm{gkp}}\rangle &=  \sum_{n\in\mathbb{Z}} |\hat{p} = (2n+1)\sqrt{\pi} \rangle. 
\end{align}
Clearly, all these basis states have $\hat{q} = \hat{p} =0$ modulo $\sqrt{\pi}$ and thus are stabilized by $\hat{S}_{q}$ and $\hat{S}_{p}$.   

The ideal GKP qubit states consist of infinitely many infinitely-squeezed states and thus are unrealistic. Realistic GKP qubit states can be obtained by applying a Gaussian envelope operator $\exp[-\Delta \hat{n}]$ to the ideal GKP states, i.e., $|\psi^{\Delta}_{\textrm{gkp}}\rangle \propto \exp[-\Delta \hat{n}] |\psi_{\textrm{gkp}}\rangle$ and have a finite average photon number or finite squeezing. Here, $\hat{n}=\hat{a}^{\dagger}\hat{a}$ is the excitation number operator and $\Delta$ characterizes the width of each peak in the Wigner function of a realistic GKP state. In \cref{fig:GKP qubit fundamentals} (a), we plot the Wigner functions of the basis states of an approximate GKP qubit with $\bar{n}=5$. There are many proposals for realizing approximate GKP states in various experimental platforms \cite{Gottesman2001,Travaglione2002,Pirandola2004,Pirandola2006,Vasconcelos2010,Terhal2016,
Motes2017,Weigand2018,Arrazola2019,Su2019,Eaton2019,Shi2019,Weigand2019}. Notably, approximate GKP states have been realized experimentally in circuit QED \cite{Campagne2019} and trapped ion systems \cite{Fluhmann2018,Fluhmann2019,Fluhmann2019b}.  In \cref{subsection:Noise models}, we discuss the adverse effects of the finite photon number in more detail. In this subsection, we instead focus on the properties of an ideal GKP qubit.

Pauli operators of the GKP qubit are given by the square root of the stabilizers, i.e., 
\begin{align}
\hat{Z}_{\textrm{gkp}} &= (\hat{S}_{q})^{\frac{1}{2}} = \exp[ i\sqrt{\pi}\hat{q}] , 
\nonumber\\
\hat{X}_{\textrm{gkp}} &= (\hat{S}_{p})^{\frac{1}{2}} = \exp [ -i\sqrt{\pi}\hat{p}] . 
\end{align}
Indeed, one can readily check that these Pauli operators act on the computational basis states as desired: 
\begin{alignat}{2}
\hat{Z}_{\textrm{gkp}}|0_{\textrm{gkp}}\rangle &= |0_{\textrm{gkp}}\rangle, \quad & \hat{Z}_{\textrm{gkp}}|1_{\textrm{gkp}}\rangle &= -|1_{\textrm{gkp}}\rangle,
\nonumber\\
\hat{X}_{\textrm{gkp}}|0_{\textrm{gkp}}\rangle &= |1_{\textrm{gkp}}\rangle, \quad & \hat{X}_{\textrm{gkp}}|1_{\textrm{gkp}}\rangle &= |0_{\textrm{gkp}}\rangle. 
\end{alignat}
Clifford operations \cite{Gottesman1999} on the GKP qubits can be implemented by using only Gaussian operations. More explicitly, generators of the Clifford group, $\hat{S}_{\textrm{gkp}}, \hat{H}_{\textrm{gkp}}$ and $\textrm{CNOT}_{\textrm{gkp}}^{j\rightarrow k}$ are given by 
\begin{align}
\hat{S}_{\textrm{gkp}} &= \exp\Big{[} i\frac{\hat{q}^{2}}{2} \Big{]} , 
\nonumber\\
\hat{H}_{\textrm{gkp}} &= \exp\Big{[} i \frac{\pi}{2}\hat{a}^{\dagger}\hat{a} \Big{]}  , 
\nonumber\\
\textrm{CNOT}_{\textrm{gkp}}^{j\rightarrow k} &= \textrm{SUM}_{j\rightarrow k} \equiv  \exp[ - i \hat{q}_{j}\hat{p}_{k}], \label{eq:Clifford gates for GKP qubits}
\end{align} 
and one can similarly check that 
\begin{alignat}{2}
\hat{S}_{\textrm{gkp}}|0_{\textrm{gkp}}\rangle &= |0_{\textrm{gkp}}\rangle, \quad & \hat{S}_{\textrm{gkp}}|1_{\textrm{gkp}}\rangle &= i|1_{\textrm{gkp}}\rangle,
\nonumber\\
\hat{H}_{\textrm{gkp}}|0_{\textrm{gkp}}\rangle &= |+_{\textrm{gkp}}\rangle, \quad & \hat{H}_{\textrm{gkp}}|1_{\textrm{gkp}}\rangle &= |-_{\textrm{gkp}}\rangle, 
\end{alignat}
and
\begin{align}
\textrm{CNOT}_{\textrm{gkp}}^{j\rightarrow k}|\mu^{(j)}_{\textrm{gkp}}\rangle|\nu^{(k)}_{\textrm{gkp}}\rangle = |\mu^{(j)}_{\textrm{gkp}}\rangle|(\mu\oplus\nu)^{(k)}_{\textrm{gkp}}\rangle, 
\end{align}
for all $\mu,\nu\in\mathbb{Z}_{2}$, where $|\mu^{(j)}_{\textrm{gkp}}\rangle \equiv \sum_{n\in\mathbb{Z}} |\hat{q}_{j} = (2n+\mu)\sqrt{\pi} \rangle$ is the GKP state in the $j^{\textrm{th}}$ mode and $\oplus$ is the addition modulo $2$.

Recall that measuring the stabilizers of the GKP qubit $\hat{S}_{q}$ and $\hat{S}_{p}$ is equivalent to measuring the position and the momentum operators $\hat{q}$ and $\hat{p}$ modulo $\sqrt{\pi}$. These measurements can be respectively performed by preparing an ancilla GKP state $|+_{\textrm{gkp}}\rangle$ or $|0_{\textrm{gkp}}\rangle$, and then applying the $\textrm{SUM}_{D \rightarrow A}$ or $\textrm{SUM}_{A\rightarrow D}^{\dagger}$ gate, and finally measuring the position or the momentum operator of the ancilla mode via a homodyne detection (see \cref{fig:GKP qubit fundamentals} (b)). Here, $D$ refers to the data mode and $A$ refers to the ancilla mode. Note that the only non-Gaussian resources required for the GKP-stabilizer measurements are the ancilla GKP states $|0_{\textrm{gkp}}\rangle$ and $|+_{\textrm{gkp}}\rangle$.    

Now, consider the Gaussian random displacement error channel $\mathcal{N}[\sigma]$ defined as 
\begin{align}
\mathcal{N}[\sigma](\hat{\rho}) \equiv \int \frac{d^{2}\alpha}{\pi \sigma^{2} }  \exp\Big{[} -\frac{|\alpha|^{2}}{\sigma^{2}} \Big{]} \hat{D}(\alpha) \hat{\rho}\hat{D}^{\dagger}(\alpha), \label{eq:Gaussian random displacement error definition}
\end{align}
where $\hat{D}(\alpha)\equiv \exp[\alpha\hat{a}^{\dagger}-\alpha^{*}\hat{a}]$ is the displacement operator and $\alpha\in\mathbb{C}$ is the amplitude of the displacement. In the Heisenberg picture, the error channel $\mathcal{N}[\sigma]$ adds shift errors to the position and momentum quadratures, that is, $\hat{q}\rightarrow \hat{q}+\xi_{q}$ and $\hat{p}\rightarrow \hat{p}+\xi_{p}$, where $\xi_{q}$ and $\xi_{p}$ follow a Gaussian random distribution with zero mean and standard deviation $\sigma$: $\xi_{q},\xi_{p}\sim \mathcal{N}(0,\sigma)$. If, for example, the size of the random position shift $\xi_{q}$ is smaller than $\sqrt{\pi}/2$ (i.e., $|\xi_{q}|< \sqrt{\pi}/2$), the shift can be correctly identified by measuring the GKP stabilizer $\hat{S}_{q}$. However, if $\xi_{q}$ lies in the range $|\xi_{q}-\sqrt{\pi}| < \sqrt{\pi}/2$, the shift is incorrectly identified as a smaller shift $\xi_{q}-\sqrt{\pi}$. Then, such a misidentification results in a residual shift $\exp[ -i\sqrt{\pi}\hat{p} ] = \hat{X}_{\textrm{gkp}}$ and thus causes a Pauli $X$ error on the GKP qubit.

In general, if $\xi_{q}$ (or $\xi_{p}$) lies in the range $|\xi_{q}-n\sqrt{\pi}|< \sqrt{\pi}/2$ (or $|\xi_{p}-n\sqrt{\pi}|< \sqrt{\pi}/2$) for an odd integer $n$, the GKP error correction protocol results in a Pauli $X$ (or $Z$) error on the GKP qubit and this happens with probability $p_{\textrm{err}}(\sigma)$, where $p_{\textrm{err}}(\sigma)$ is defined as 
\begin{align}
p_{\textrm{err}}(\sigma) &\equiv  \sum_{n\in\mathbb{Z}} \frac{1}{\sqrt{2\pi\sigma^{2}}} \int_{(2n+\frac{1}{2} ) \sqrt{\pi} }^{ (2n+\frac{3}{2})\sqrt{\pi} } d\xi  \exp\Big{[} -\frac{\xi^{2}}{2\sigma^{2}} \Big{]}. \label{eq:definition of perr}
\end{align}
Now, consider a specific instance where, for example, the $\hat{S}_{q}$ stabilizer measurement (i.e., the position measurement modulo $\sqrt{\pi}$) informs us that $\xi_{q}$ is given by $\xi_{q} = z + n\sqrt{\pi}$ for some interger $n$ and $|z| < \sqrt{\pi}/2$. Then, since odd $n$ corresponds to a Pauli $X$ error and even $n$ corresponds to the no error case, we can infer that, given the measured value $z$, there is a Pauli $X$ error with probability $p[\sigma](z)$ where $p[\sigma](z)$ is defined as  
\begin{align}
p[\sigma](z) &\equiv \frac{ \sum_{n\in\mathbb{Z}} \exp[- (z-(2n+1)\sqrt{\pi})^{2}   / (2\sigma^{2})  ] }{ \sum_{n\in\mathbb{Z}} \exp[- (z-n\sqrt{\pi})^{2}  / (2\sigma^{2}) ]  }.  \label{eq:Definition of the p function}
\end{align}

\begin{figure}[t!]
\centering
\includegraphics[width=3.2in]{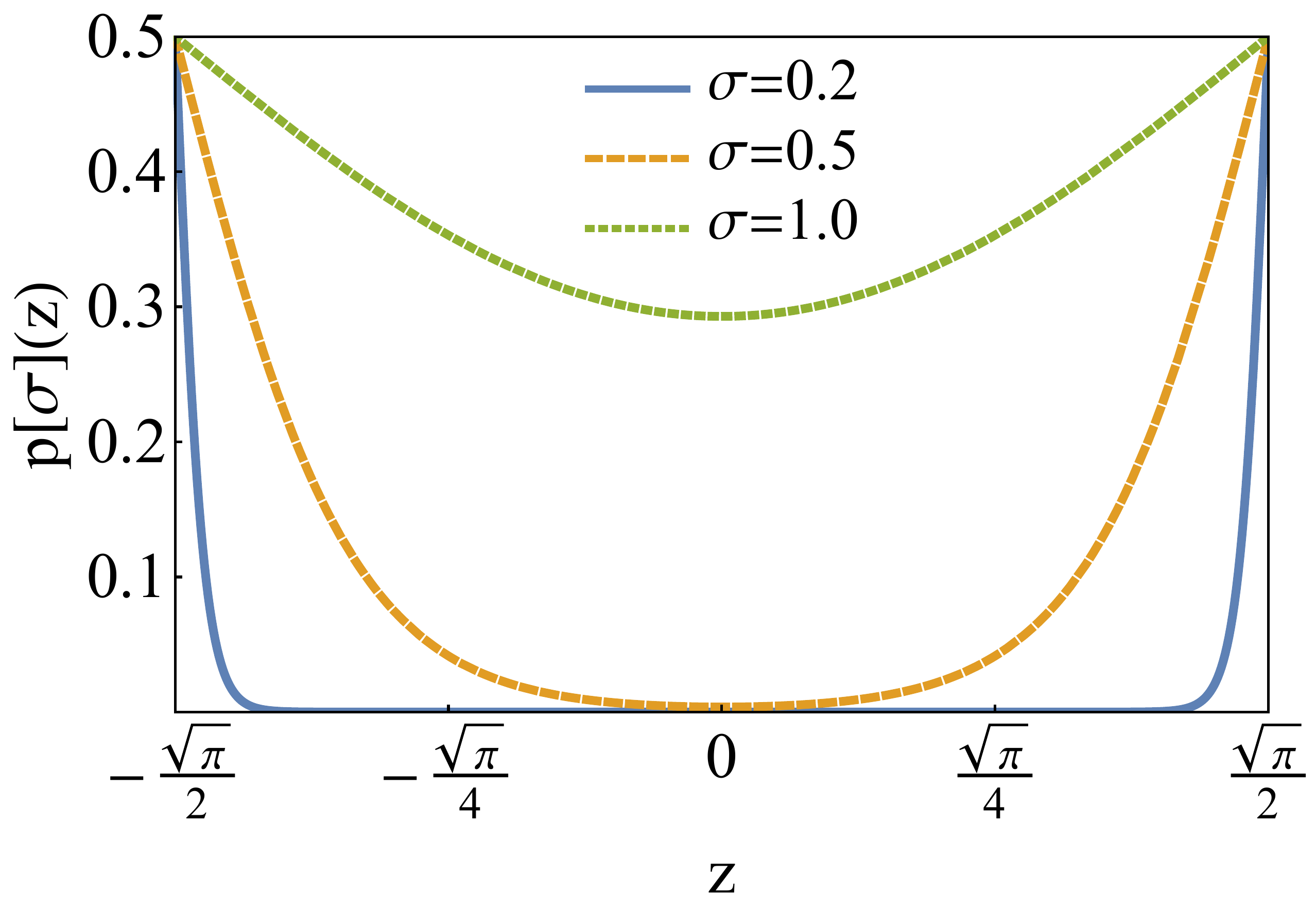}
\caption{$p[\sigma](z)$ for $\sigma = 0.2$, $0.5$, and $1$. $p[\sigma](z)$ is defined in \cref{eq:Definition of the p function} and represents the conditional probability of having a Pauli $X$ (or $Z$) error, given the measurement outcome $\xi_{q}=z+n\sqrt{\pi}$ (or $\xi_{p}=z+n\sqrt{\pi}$) for some integer $n$.  }
\label{fig:conditional probability p}
\end{figure}

As shown in \cref{fig:conditional probability p}, the conditional probability $p[\sigma](z)$ becomes larger as $|z|$ gets closer to the decision boundary $\sqrt{\pi}/2$. Therefore, if the measured shift value modulo $\sqrt{\pi}$ is close to $\pm\sqrt{\pi}/2$, we know that this specific instance of the GKP error correction is less reliable. This way, the GKP error correction protocol not only corrects the small shift errors but also informs us how reliable the correction is. Various ways of incorporating this additional information in the next level of concatenated error correction have been studied in Refs. \cite{Fukui2017,Fukui2018a,Fukui2018b,Fukui2019,Vuillot2019}. In \cref{section:Methods}, we explain in detail how the additional information from GKP-stabilizer measurements can be used to compute renormalized edge weights of the matching graphs used in the surface code error correction protocol.

Lastly, although not relevant to the purpose of our work, it has been shown that a H-type GKP-magic state $|H_{\textrm{gkp}}\rangle = \cos (\frac{\pi}{8}) |0_{\textrm{gkp}}\rangle + \sin (\frac{\pi}{8}) |1_{\textrm{gkp}}\rangle $ can be prepared by performing GKP-stabilizer measurements on a vacuum state and then post-selecting the $\hat{S}_{q} = \hat{S}_{p} = 1$ (or $\hat{q} = \hat{p} =0$ modulo $\sqrt{\pi}$) event \cite{Terhal2016} (see Ref. \cite{Bravyi2005} for more details on the magic states). Notably, a more recent study \cite{Baragiola2019} has quantitatively showed that any post-measurement state after the GKP-stabilizer measurements (on a vacuum state) is a distillable GKP-magic state and therefore post-selection is not necessary. Since Clifford operations (necessary for magic state distillation) on GKP qubits can be implemented by using only Gaussian operations, the ability to prepare GKP states is the only non-Gaussian resource needed for universal quantum computation using GKP qubits.

\begin{figure*}[t!]
\centering
\includegraphics[width=6.8in]{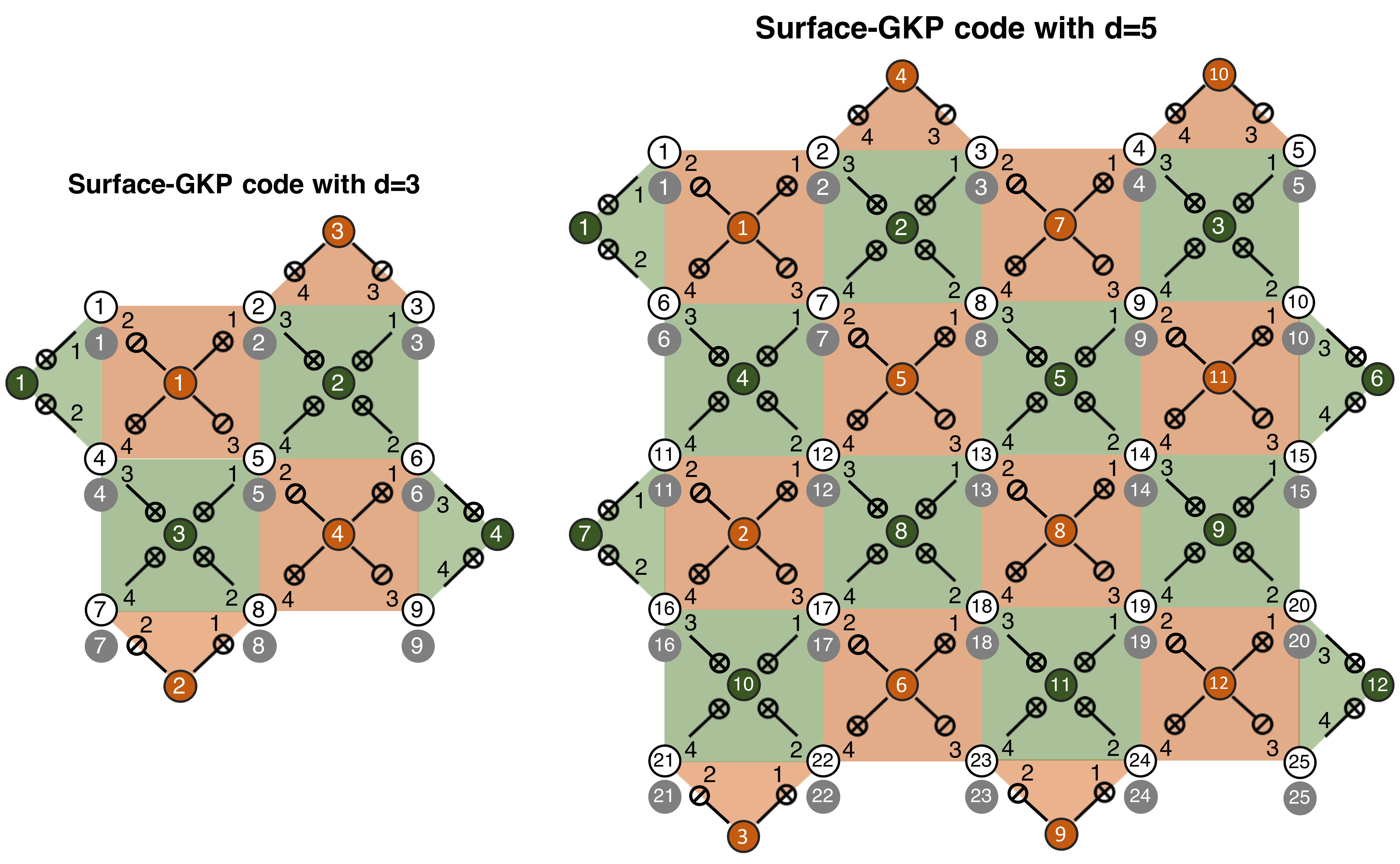}
\caption{The surface-GKP codes with $d=3$ and $d=5$. White circles represent the data GKP qubits and grey circles represent the ancilla GKP qubits that are used to measure GKP stabilizers of each data GKP qubit. Green and orange circles represent the syndrome GKP qubits that are used to measure the $Z$-type and $X$-type surface code stabilizers of the data GKP qubits, respectively. In general, there are $d^{2}$ data GKP qubits and $(d^{2}-1)/2$ $Z$-type and $X$-type syndrome GKP qubits. See also \cref{fig:Surface code error propagation} for the reason behind our choice of inverse-SUM gates in the $X$-type stabilizer measurements.    }
\label{fig:Surface-GKP codes}
\end{figure*}

\subsection{The surface code with GKP qubits}
\label{subsection:Surface codes with GKP qubits}

\begin{figure}[b!]
\centering
\includegraphics[width=3.35in]{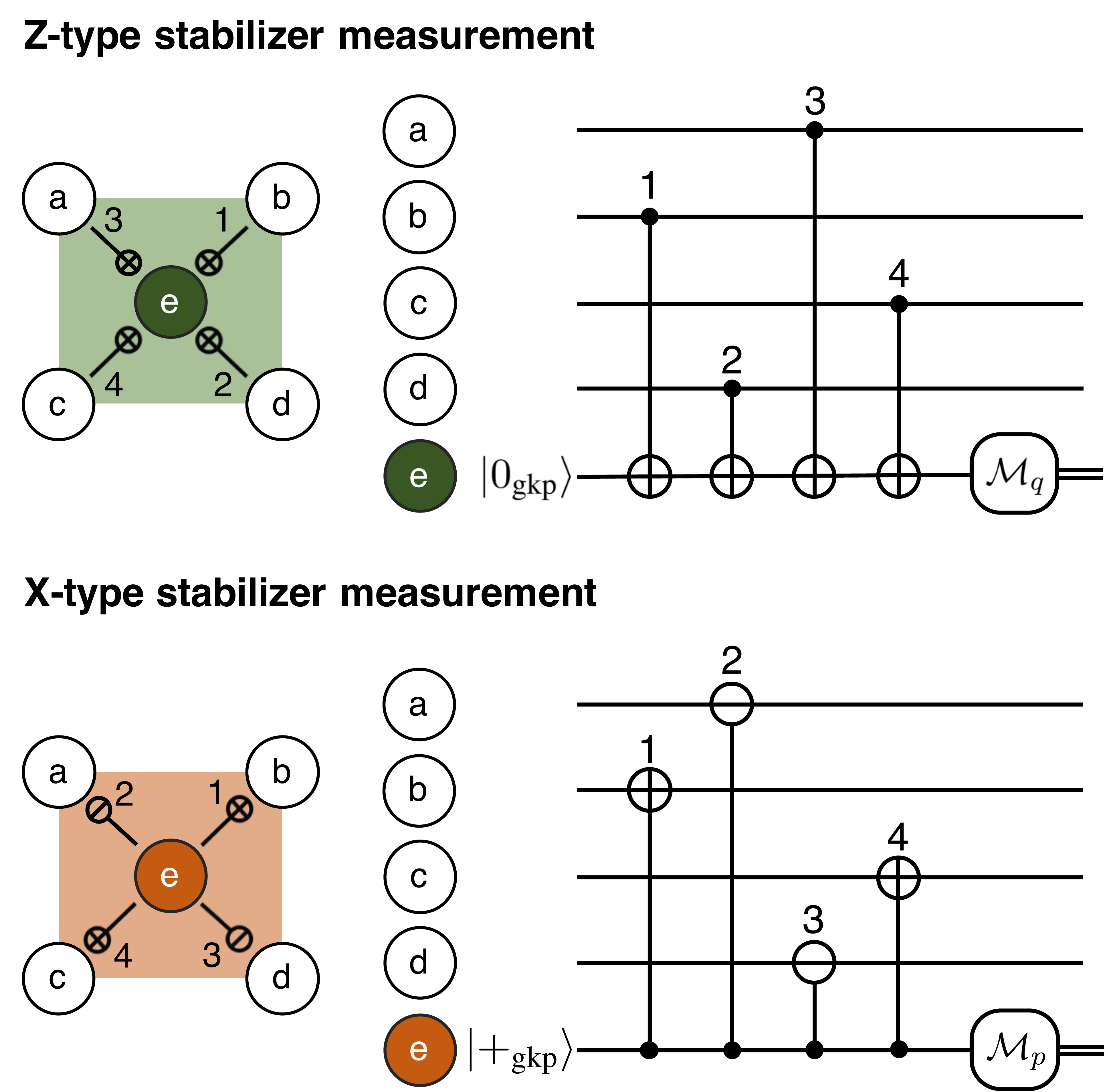}
\caption{Circuits for surface code stabilizer measurements.   }
\label{fig:Surface code stabilizer measurement}
\end{figure}

Recall that shift errors of size larger than $\sqrt{\pi}/2$ cannot be corrected by the single-mode GKP code. Here, to correct arbitrarily large shift errors, we consider the concatenation of the GKP code with the surface code \cite{Bravyi1998,Dennis2002,Fowler2012}, namely, the surface-GKP code. Specifically, we use the family of rotated surface codes \cite{Bombin2007,Tomita2014} that only requires $d^{2}$ data qubits and $d^{2}-1$ syndrome qubits to get a distance-$d$ code. Note that the distance-$d$ surface code can correct arbitrary qubit errors of weight less than or equal to $\lfloor \frac{d-1}{2} \rfloor$.

\begin{figure*}[t!]
\centering
\includegraphics[width=5.4in]{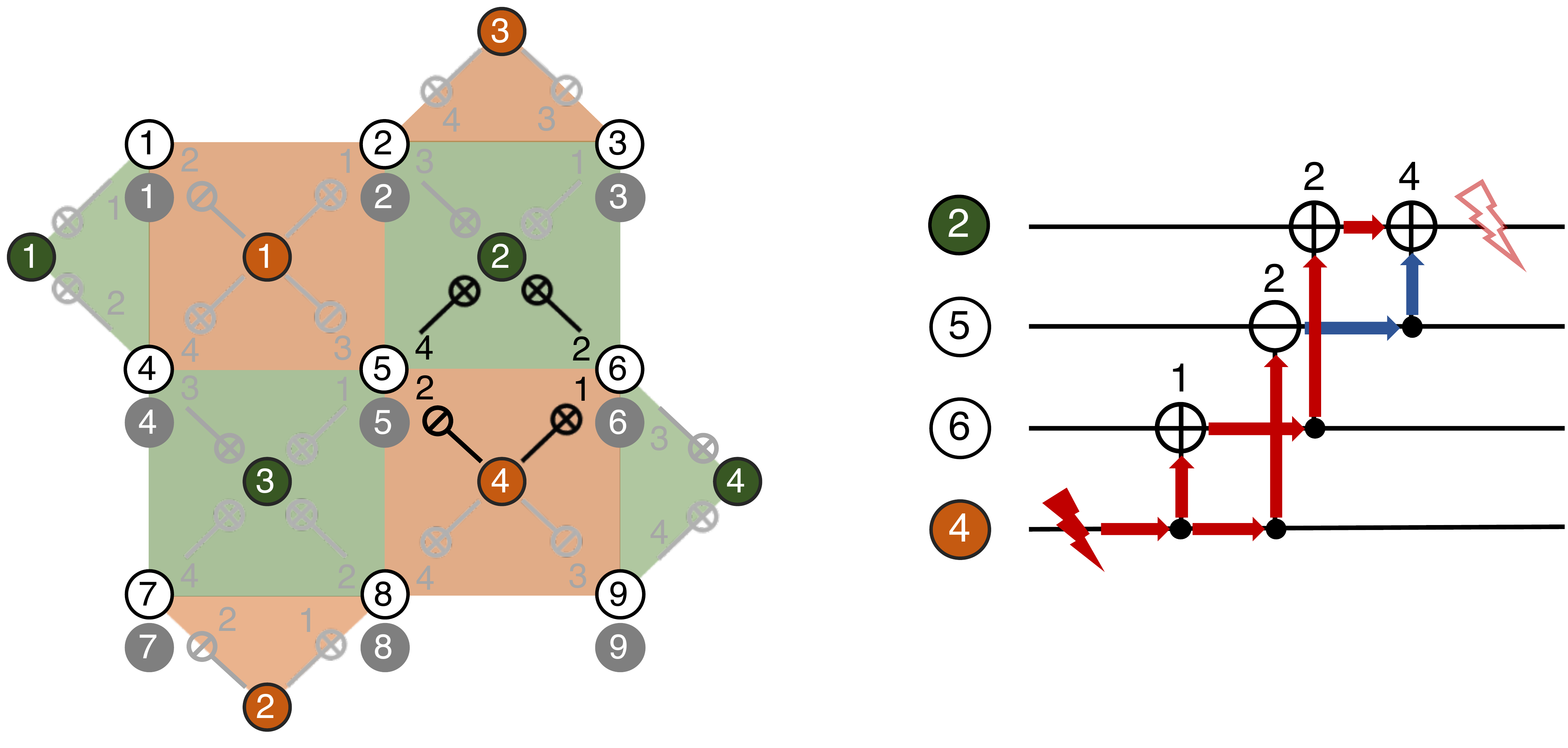}
\caption{Noise propagation from the X4 qubit to the Z2 qubit during surface code stabilizer measurements. The red lightening symbol represents the initial position of a shift error on the qubit X4. During the propagation of the shift error to the qubit Z2, the sign of the shift error is flipped by the inverse-SUM gate $\textrm{SUM}^{\dagger}_{X4\rightarrow D5}$. This sign flip then results in cancellations of the propagated shift errors on the qubit Z2 (empty lightening symbol).       }
\label{fig:Surface code error propagation}
\end{figure*}

The layout for the data and ancilla qubits of the surface-GKP code is given in \cref{fig:Surface-GKP codes}. Each of the $d^{2}$ data qubits (white circles in \cref{fig:Surface-GKP codes}) corresponds to a GKP qubit as defined in \cref{subsection:GKP qubit}. That is, the distance-$d$ surface-GKP code is stabilized by the following $2d^{2}$ GKP stabilizers 
\begin{align}
\hat{S}_{q}^{(k)} \equiv \exp[ i2\sqrt{\pi}\hat{q}_{k} ], \quad \hat{S}_{p}^{(k)} \equiv \exp[ -i2\sqrt{\pi}\hat{p}_{k} ], 
\end{align}
for $k\in \lbrace 1,\cdots, d^{2}  \rbrace$. These GKP stabilizers are measured by $d^{2}$ ancilla GKP qubits (grey circles in \cref{fig:Surface-GKP codes}) using the circuits given in \cref{fig:GKP qubit fundamentals} (b). Moreover, the data GKP qubits are further stabilized by the $d^{2}-1$ surface code stabilizers. For example, in the $d=3$ case, the $8$ surface code stabilizers are explicitly given by    
\begin{alignat}{2}
\hat{S}_{Z}^{(1)} &= \hat{Z}_{\textrm{gkp}}^{(1)}\hat{Z}_{\textrm{gkp}}^{(4)},\quad & \hat{S}_{Z}^{(2)} &= \hat{Z}_{\textrm{gkp}}^{(2)}\hat{Z}_{\textrm{gkp}}^{(3)}\hat{Z}_{\textrm{gkp}}^{(5)}\hat{Z}_{\textrm{gkp}}^{(6)}, 
\nonumber\\
\hat{S}_{Z}^{(3)} &= \hat{Z}_{\textrm{gkp}}^{(4)}\hat{Z}_{\textrm{gkp}}^{(5)}\hat{Z}_{\textrm{gkp}}^{(7)}\hat{Z}_{\textrm{gkp}}^{(8)}, \quad & \hat{S}_{Z}^{(4)} &= \hat{Z}_{\textrm{gkp}}^{(6)}\hat{Z}_{\textrm{gkp}}^{(9)}, \label{eq:surface code stabilizers Z type d=3}
\end{alignat}
and
\begin{align}
\hat{S}_{X}^{(1)} &= (\hat{X}_{\textrm{gkp}}^{(1)})^{\dagger}\hat{X}_{\textrm{gkp}}^{(2)}\hat{X}_{\textrm{gkp}}^{(4)}(\hat{X}_{\textrm{gkp}}^{(5)})^{\dagger},\quad  \hat{S}_{X}^{(2)} = (\hat{X}_{\textrm{gkp}}^{(7)})^{\dagger}\hat{X}_{\textrm{gkp}}^{(8)},  
\nonumber\\
\hat{S}_{X}^{(3)} &= \hat{X}_{\textrm{gkp}}^{(2)}(\hat{X}_{\textrm{gkp}}^{(3)})^{\dagger}, \quad  \hat{S}_{X}^{(4)} = (\hat{X}_{\textrm{gkp}}^{(5)})^{\dagger}\hat{X}_{\textrm{gkp}}^{(6)}\hat{X}_{\textrm{gkp}}^{(8)}(\hat{X}_{\textrm{gkp}}^{(9)})^{\dagger}, \label{eq:surface code stabilizers X type d=3}
\end{align}
where $\hat{Z}_{\textrm{gkp}}^{(k)} \equiv \exp[i\sqrt{\pi}\hat{q}_{k}]$ and $\hat{X}_{\textrm{gkp}}^{(k)} \equiv \exp[-i\sqrt{\pi}\hat{p}_{k}]$ (see \cref{fig:Surface-GKP codes}).

As shown in \cref{fig:Surface code stabilizer measurement}, the $Z$-type surface code stabilizers are measured by the $Z$-type GKP syndrome qubits (green circles in \cref{fig:Surface-GKP codes}) by using the SUM gates $\textrm{SUM}_{a\rightarrow e},\cdots, \textrm{SUM}_{d\rightarrow e}$ and the position homodyne measurement $\mathcal{M}_{q}$. Similarly, the $X$-type surface code stabilizers are measured by the $X$-type GKP syndrome qubits (orange circles in \cref{fig:Surface-GKP codes}) by using the SUM and the inverse-SUM gates $\textrm{SUM}_{e\rightarrow a}^{\dagger},\textrm{SUM}_{e\rightarrow b},\textrm{SUM}_{e\rightarrow c},\textrm{SUM}_{e\rightarrow d}^{\dagger}$ and the momentum homodyne measurement $\mathcal{M}_{p}$. Note that all the $Z$-type and $X$-type surface code stabilizers can be measured in parallel without conflicting with each other, if the SUM and the inverse-SUM gates are executed in an order that is specified in \cref{fig:Surface-GKP codes,fig:Surface code stabilizer measurement}.

We remark that in the usual case where the surface code is implemented with bare qubits (such as transmons \cite{Koch2007,Schreier2008}), it makes no difference to replace, for example, $\hat{S}_{X}^{(1)} = (\hat{X}^{(1)})^{\dagger}\hat{X}^{(2)}\hat{X}^{(4)}(\hat{X}^{(5)})^{\dagger}$ by $\hat{S}_{X}^{(1)} = \hat{X}^{(1)}\hat{X}^{(2)}\hat{X}^{(4)}\hat{X}^{(5)}$ since the Pauli operators are hermitian. Similarly, the action of $(\hat{X}_{\textrm{gkp}}^{(k)})^{\dagger}$ on the GKP qubit subspace is identical to that of $\hat{X}_{\textrm{gkp}}^{(k)}$ and therefore measuring $\hat{S}_{X}^{(1)} = (\hat{X}_{\textrm{gkp}}^{(1)})^{\dagger}\hat{X}_{\textrm{gkp}}^{(2)}\hat{X}_{\textrm{gkp}}^{(4)}(\hat{X}_{\textrm{gkp}}^{(5)})^{\dagger}$ is equivalent to measuring $\hat{S}_{X}^{(1)} = \hat{X}_{\textrm{gkp}}^{(1)}\hat{X}_{\textrm{gkp}}^{(2)}\hat{X}_{\textrm{gkp}}^{(4)}\hat{X}_{\textrm{gkp}}^{(5)}$ in the case of the surface-GKP code if the syndrome measurements are noiseless.

It is important to note, however, that the actions of $(\hat{X}_{\textrm{gkp}}^{(k)})^{\dagger}$ and $\hat{X}_{\textrm{gkp}}^{(k)}$ are not the same \textit{outside} of the GKP qubit subspace. Therefore, it does make a difference to choose $(\hat{X}_{\textrm{gkp}}^{(k)})^{\dagger}$ instead of $\hat{X}_{\textrm{gkp}}^{(k)}$ in the noisy measurement case, since shift errors propagate differently depending on the choice. For example, we illustrate in \cref{fig:Surface code error propagation} how the initial position shift error in the fourth $X$-type syndrome GKP qubit (X4 qubit) propagates to the second $Z$-type syndrome GKP qubit (Z2 qubit) through the fifth and the sixth data GKP qubits (D5 and D6 	qubits). Note that an initial random position shift in the X4 qubit (represented by the red lightning symbol) is propagated to the D6 qubit via the SUM gate $\textrm{SUM}_{X4\rightarrow D6}$ and then to the Z2 qubit via $\textrm{SUM}_{D6\rightarrow Z2}$. Additionally, it is also propagated to the D5 qubit via the inverse-SUM gate $\textrm{SUM}^{\dagger}_{X4\rightarrow D5}$ with its sign flipped and then the flipped shift is further propagated to the Z2 qubit via $\textrm{SUM}_{D5\rightarrow Z2}$. Thus, the propagated shift errors eventually cancel out each other at the Z2 qubit (visualized by the empty lightning symbol) due to the sign flip during the inverse-SUM gate. 

Note that if the SUM gate $\textrm{SUM}_{X4\rightarrow D5}$ were used instead of the inverse-SUM gate $\textrm{SUM}^{\dagger}_{X4\rightarrow D5}$, the propagated shift errors would add together and therefore be amplified by a factor of $2$. In this regard, we emphasize that we have carefully chosen the specific pattern of the SUM and the inverse-SUM gates in \cref{fig:Surface-GKP codes} to avoid such noise amplifications.

\subsection{Noise model} 
\label{subsection:Noise models}

In this section we discuss the noise model that we use to simulate the full error correction protocol with the surface-GKP code. To be more specific, the surface-GKP error correction protocol is implemented by repeatingly measuring the $\hat{S}_{q}$ and $\hat{S}_{p}$ GKP stabilizers for each data GKP qubit by using the circuits in \cref{fig:GKP qubit fundamentals} (b), and then measuring the surface code stabilizers shown in \cref{fig:Surface-GKP codes,fig:Surface code stabilizer measurement}. Note that the required resources for these measurements are as follows:  
\begin{itemize}
\item Preparation of the GKP states $|0_{\textrm{gkp}}\rangle$ and $|+_{\textrm{gkp}}\rangle$. 
\item SUM and inverse-SUM gates.
\item Position and momentum homodyne measurements.
\item Displacement operations for error correction.
\end{itemize}
We assume that all these components can be noisy except for the displacement operations since in most experimental platforms, the errors associated with the displacement operations are negligible compared to the other errors. Moreover, note that displacement operations are only needed for error correction. Thus, they need not be implemented physically in practice since they can be kept track of by using a Pauli frame \cite{Knill2005, DiVincenzo2007,Terhal2015,Chamberland2018}. Below, we describe the noise model for each component in more detail. 

Let us recall that realistic GKP states have a finite average photon number, or finite squeezing. As discussed in \cref{subsection:GKP qubit}, a finite-size GKP state can be modeled by applying a Gaussian envelope operator $\exp[-\Delta\hat{n}]$ to an ideal GKP state, i.e., $|\psi_{\textrm{gkp}}^{\Delta}\rangle\propto \exp[-\Delta\hat{n}]|\psi_{\textrm{gkp}}\rangle$. Expanding the envelope operator in terms of displacement operators \cite{Cahill1969}, we can write 
\begin{align}
|\psi_{\textrm{gkp}}^{\Delta}\rangle &\propto  \int  \frac{d^{2}\alpha}{\pi}  \textrm{Tr}\big{[} \exp[-\Delta\hat{n}] \hat{D}^{\dagger}(\alpha)\big{]} \hat{D}(\alpha) |\psi_{\textrm{gkp}} \rangle 
\nonumber\\
&\propto \int d^{2}\alpha \exp\Big{[} -\frac{|\alpha|^{2}}{ 2\sigma_{\textrm{gkp}}^{2} }  \Big{]} \hat{D}(\alpha) |\psi_{\textrm{gkp}} \rangle , \label{eq:finite GKP expansion in displacements}
\end{align} 
where $\sigma_{\textrm{gkp}}^{2} = (1-e^{-\Delta}) / (1+e^{-\Delta}) \xrightarrow{\Delta\ll 1} \Delta /2$ (see \cref{appeq:laguerre identity}). That is, an approximate GKP state can be understood as the state that results from applying coherent superpositions of displacement operations with a Gaussian envelope to an ideal GKP state. More details about the approximate GKP codes can be found in \cite{Terhal2016,Shi2019,Pantaleoni2019,Tzitrin2019,Matsuura2019}. 

To simplify our analysis of the surface-GKP code, we consider noisy GKP states corrupted by an \textit{incoherent} mixture of displacement operations, instead of the coherent superposition as in \cref{eq:finite GKP expansion in displacements}. That is, whenever a fresh GKP state $|0_{\textrm{gkp}}\rangle$ or $|+_{\textrm{gkp}}\rangle$ is supplied to the error correction chain, we assume that a noisy GKP state
\begin{align}
|0_{\textrm{gkp} } \rangle &\rightarrow \mathcal{N}[\sigma_{\textrm{gkp}}](|0_{\textrm{gkp} } \rangle\langle 0_{\textrm{gkp}} | ), \textrm{ or}
\nonumber\\
|+_{\textrm{gkp} } \rangle &\rightarrow \mathcal{N}[\sigma_{\textrm{gkp}}](|+_{\textrm{gkp} } \rangle\langle +_{\textrm{gkp}} | ) \label{eq:noisy GKP state incoherent displacement error}
\end{align}
is supplied where the Gaussian random displacement error $\mathcal{N}[\sigma]$ is defined in \cref{eq:Gaussian random displacement error definition}. Note that $\mathcal{N}[\sigma]$ models an incoherent mixture of random displacement errors. We remark that the noisy GKP states corrupted by an incoherent displacement error (as in \cref{eq:noisy GKP state incoherent displacement error}) are noisier than the noisy GKP states corrupted by a coherent displacement error (as in \cref{eq:finite GKP expansion in displacements}), because the former can be obtained from the latter by applying a technique similar to Pauli twirling \cite{Emerson2007} (see \cref{appendix:Supplementary material for noisy GKP states}). In this sense, by adopting the incoherent noise model, we make a conservative assumption about the GKP noise while simplifying the analysis. 

We define the squeezing $s_{\textrm{gkp}}$ of a noisy GKP state $\mathcal{N}[\sigma_{\textrm{gkp}}](|\psi_{\textrm{gkp} } \rangle\langle \psi_{\textrm{gkp}} | )$ as $s_{\textrm{gkp}}\equiv -10\log_{10}(2\sigma_{\textrm{gkp}}^{2})$ (aligning our notation with those in Refs. \cite{Menicucci2014,Fukui2018a,Fukui2019}), where the unit of $s_{\textrm{gkp}}$ is in dB. We also assume that idling modes are undergoing independent Gaussian random displacement errors $\mathcal{N}[\sigma_{p}]$ with variance $\sigma_{p}^{2} = \kappa \Delta t_{p}$ during the GKP state preparation, where $\kappa$ is the photon loss and heating rate (see below) and $\Delta t_{p}$ is the time needed to prepare the GKP states. 

Secondly, we assume that photon loss errors occur continuously during the execution of the SUM or the inverse-SUM gates. To be more specific, we assume that SUM gates are implemented by letting the system evolve under the Hamiltonian $\hat{H} = g\hat{q}_{1}\hat{p}_{2}$ for $\Delta t = 1/g$ (the first mode is the control mode and the second mode is the target mode), during which independent photon loss errors occur continuously in both the control and the target mode. That is, we replace the unitary SUM gate $\textrm{SUM}_{1\rightarrow 2} = \exp[-i\hat{q}_{1}\hat{p}_{2}]$ (or the inverse-SUM gate $\textrm{SUM}_{1\rightarrow 2}^{\dagger} = \exp[i\hat{q}_{1}\hat{p}_{2}]$) by a completely positive and trace-preserving (CPTP) map \cite{Choi1975} $\exp[\mathcal{L}_{+}\Delta t]$ (or $\exp[\mathcal{L}_{-}\Delta t]$) with $\Delta t = 1/g$, where $g$ is the coupling strength and the Lindbladian generator $\mathcal{L}_{\pm}$ is given by 
\begin{align}
\mathcal{L}_{\pm }(\hat{\rho}) = \mp ig[ \hat{q}_{1}\hat{p}_{2}, \hat{\rho}  ] + \kappa \big{(} \mathcal{D}[\hat{a}_{1}] + \mathcal{D}[\hat{a}_{2}] \big{)}\hat{\rho}. \label{eq:noisy SUM or inverse-SUM gates Lindbladian}
\end{align}
Here, $\mathcal{D}[\hat{A}](\hat{\rho}) \equiv \hat{A}\hat{\rho}\hat{A}^{\dagger} - \frac{1}{2}\lbrace \hat{A}^{\dagger}\hat{A},\hat{\rho} \rbrace$, and $\kappa$ is the photon loss rate. 

In a similar spirit as above, we make a more conservative assumption about the gate error to make the analysis more tractable. That is, we make the noisy gate $\exp[\mathcal{L}_{\pm}\Delta t]$ noisier by adding heating errors $\kappa( \mathcal{D}[\hat{a}_{1}^{\dagger}] + \mathcal{D}[\hat{a}_{2}^{\dagger}] )$ to the Lindbladian $\mathcal{L}_{\pm}$, i.e., 
\begin{align}
\mathcal{L}'_{\pm} \equiv  \mathcal{L}_{\pm} + \kappa\big{(} \mathcal{D}[\hat{a}_{1}^{\dagger}] + \mathcal{D}[\hat{a}_{2}^{\dagger}] \big{)} , 
\end{align}
where the heating rate $\kappa$ is the same as the photon loss rate. This is to convert the loss errors into random displacement errors (see Refs. \cite{Albert2018,Noh2019}). Indeed, the noisy SUM or the inverse-SUM gate $\exp[\mathcal{L}'_{\pm}\Delta t]$ is equivalent to the ideal SUM or the inverse-SUM gate followed by a correlated Gaussian random displacement error $\hat{q}_{k}\rightarrow \hat{q}_{k} + \xi_{q}^{(k)}$ and $\hat{p}_{k}\rightarrow \hat{p}_{k} + \xi_{p}^{(k)}$ for $k\in \lbrace 1,2 \rbrace$, where the additive shift errors are drawn from bivariate Gaussian distributions $(\xi_{q}^{(1)},\xi_{q}^{(2)}) \sim \mathcal{N}(0,\boldsymbol{N}_{q}^{\pm})$ and $(\xi_{p}^{(1)},\xi_{p}^{(2)}) \sim \mathcal{N}(0,\boldsymbol{N}_{p}^{\pm})$ with the noise covariance matrices 
\begin{align}
\boldsymbol{N}_{q}^{\pm} = \sigma_{c}^{2}\begin{bmatrix}
1 & \pm 1/2\\
\pm 1/2 & 4/3
\end{bmatrix},  \,\,  \boldsymbol{N}_{p}^{\pm} = \sigma_{c}^{2}\begin{bmatrix}
4/3 & \mp 1/2\\
\mp 1/2 & 1
\end{bmatrix}. \label{eq:noise covariance matrix SUM or inverse-SUM}
\end{align}
Here, the variance $\sigma_{c}^{2}$ is given by $\sigma_{c}^{2} = \kappa \Delta t = \kappa /g$. The noise covariance matrices $\boldsymbol{N}_{q}^{+}$ and $\boldsymbol{N}_{p}^{+}$ are used for the SUM gate and $\boldsymbol{N}_{q}^{-}$ and $\boldsymbol{N}_{p}^{-}$ are used for the inverse-SUM gate. If there are idling modes during the application of the SUM or the inverse-SUM gates on some other pairs of modes, we assume that the idling modes undergo independent Gaussian random displacement errors $\mathcal{N}[\sigma_{c}]$ of the same variance $\sigma_{c}^{2} = \kappa \Delta t = \kappa/g$, because they should wait for the same amount of time until the gates are completed.   

Lastly, we model errors in position and momentum homodyne measurements by adding independent Gaussian random displacement errors $\mathcal{N}[\sigma_{m}]$ of the variance $\sigma_{m}^{2} = \kappa \Delta t_{m}$ before the ideal homodyne measurements. Here, $\Delta t_{m}$ is the time needed to implement the homodyne measurements. Also, during the homodyne measurements, we assume that idling modes are undergoing independent Gaussian random displacement errors of the same variance $\sigma_{m}^{2} = \kappa \Delta t_{m}$.

\section{Main results}  
\label{section:Main results}

In this section, we rigorously analyze the performance of the surface-GKP code by simulating the full error correction protocol assuming the noise model described in \cref{subsection:Noise models}. We focus on the case $\sigma_{p}=\sigma_{c}=\sigma_{m}\equiv \sigma$ where all circuit elements are comparably noisy. However, we assume that the noise afflicting GKP states $\sigma_{\textrm{gkp}}$ is independent of the circuit noise. Since we have two independent noise parameters $\sigma_{\textrm{gkp}}$ and $\sigma$, the fault-tolerance thresholds would form a curve instead of a single number. Therefore, instead of exhaustively investigating the entire parameter space, we consider the following three representative scenarios:  
\begin{enumerate}[label={Case \Roman*},wide =\parindent]
\item \!\!: $\sigma_{\textrm{gkp}} \neq 0$ and $\sigma=0$ \label{case:1}
\item \!\!: $\sigma_{\textrm{gkp}} = 0$ and $\sigma\neq 0$ \label{case:2}
\item \!\!: $\sigma_{\textrm{gkp}} = \sigma \neq  0$ \label{case:3}
\end{enumerate}  
Then, we find the threshold values for $\sigma_{\textrm{gkp}}$ (\ref{case:1}), $\sigma$ (\ref{case:2}), and $\sigma_{\textrm{gkp}}=\sigma$ (\ref{case:3}), under which fault-tolerant quantum error correction is possible with the surface-GKP code. Specifically, we take the distance $d$ surface-GKP code and repeat the (noisy) stabilizer measurements $d$ times. Then, we construct 3D space-time graphs based on the stabilizer measurement outcomes and apply a minimum-weight perfect matching decoding algorithm \cite{Edmonds1965,Edmonds1965b} to perform error correction. Specifically, we use a simple method to compute the renormalized edge weights of the 3D matching graphs, based on the information obtained during GKP-stabilizer measurements. Such graphs are then used to perform MWPM.  A detailed description of our method is given in \cref{section:Methods}. Below, we report the logical $X$ error rates, which are the same as the logical $Z$ error rates. Logical $Y$ error rates are not shown since they are much smaller than the logical $X$ and $Z$ error rates.

\begin{figure*}[t!]
\centering
\includegraphics[width=5.2in]{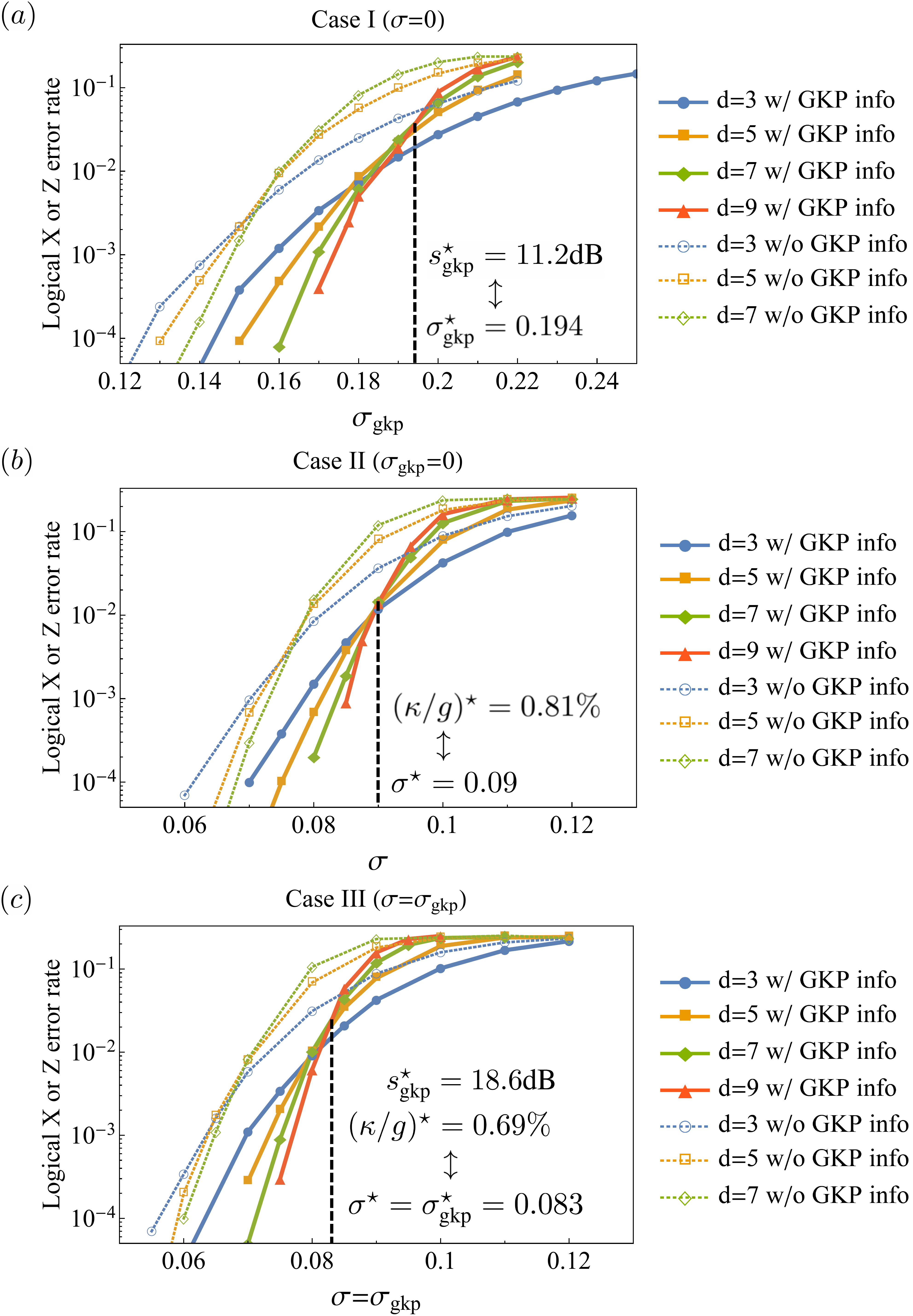}
\caption{The logical $X$ error rate of the surface-GKP code for various $d$ when (a) $\sigma=0$ (\ref{case:1}), (b) $\sigma_{\textrm{gkp}}=0$ (\ref{case:2}), and (c) $\sigma = \sigma_{\textrm{gkp}}$ (\ref{case:3}), which is the same as the logical $Z$ error rate. The solid lines represent logical error rates when information from the GKP-stabilizer measurements is used to renormalize edge weights in the matching graphs. The dotted lines correspond to the case when information from GKP-stabilizer measurements is ignored. In all cases, given that $\sigma_{\textrm{gkp}}$ and $\sigma$ are below certain fault-tolerance thresholds, the logical $X$ or $Z$ error rates are suppressed to an arbitrarily small value as we increase the code distance $d$.     }
\label{fig:main results}
\end{figure*}

In \cref{fig:main results} (a), we consider the case where GKP states are the only noisy components in the scheme, i.e., $\sigma=0$ (\ref{case:1}). We show the performance of the surface-GKP code when both the additional information from GKP-stabilizer measurements is incorporated and when it is ignored. When the additional information is incorporated, the logical $X$ error rate (same as the logical $Z$ error rate) decreases as we increase the code distance $d$ if $\sigma_{\textrm{gkp}}$ is smaller than the threshold value $\sigma_{\textrm{gkp}}^{\star} = 0.194$ (or if the squeezing of the noisy GKP state $s_{\textrm{gkp}}$ is higher than the threshold value $s_{\textrm{gkp}}^{\star} = 11.2$dB). That is, in this case, fault-tolerant error correction is possible with the surface-GKP code if the squeezing of the GKP states is above $11.2$dB. Note that if the additional information from GKP-stabilizer measurements is ignored, the threshold squeezing value decreases and logical error rates can range from one to several orders of magnitude larger for a given $\sigma_{\textrm{gkp}}$.      

In \cref{fig:main results} (b), we consider the case where GKP states are noiseless but the other circuit elements are noisy, i.e., $\sigma_{\textrm{gkp}}=0$ (\ref{case:2}). In this case, if the additional information from the GKP error correction protocol is incorporated, we can suppress the logical $X$ error rate (same as the logical $Z$ error rate) to any desired small value by choosing a sufficiently large code distance $d$ as long as $\sigma$ is smaller than the threshold value $\sigma^{\star} = 0.09$. Note that since $\sigma^{2} = \kappa / g$ the threshold value $\sigma^{\star} = 0.09$ corresponds to $(\kappa / g)^{\star} = 8.1\times 10^{-3} = 0.81\%$, where $\kappa$ is the photon loss rate and $g$ is the coupling strength of the SUM or the inverse-SUM gates. That is, fault-tolerant error correction with the surface-GKP code is possible if the SUM or the inverse-SUM gates can be implemented roughly $120$ times faster than the photon loss processes. Note that if the additional information from GKP-stabilizer measurements is ignored, the threshold value becomes smaller and logical error rates can range from one to several orders of magnitude larger for a given $\sigma$.  

Finally in \cref{fig:main results} (c), we consider the case where the GKP states and the other circuit elements are comparably noisy, i.e., $\sigma = \sigma_{\textrm{gkp}}$ (\ref{case:3}). In this case, fault-tolerant error correction is possible if $\sigma = \sigma_{\textrm{gkp}}$ is smaller than the threshold value $\sigma^{\star}=\sigma^{\star}_{\textrm{gkp}} = 0.083$. This threshold value corresponds to the GKP squeezing $s_{\textrm{gkp}}^{\star}=18.6$dB and $\kappa / g = 6.9\times 10^{-3} = 0.69\%$. Similarly, as in the previous cases, if the additional information from GKP-stabilizer measurements is ignored, the threshold value becomes smaller and logical error rates can range from one to several orders of magnitude larger for a given noise parameter $\sigma = \sigma_{\textrm{gkp}}$. 

For all three cases, we clearly observe that fault-tolerant quantum error correction with the surface-GKP code is possible despite noisy GKP states and noisy circuit elements, given that the noise parameters are below certain fault-tolerance thresholds. Recent state-of-the-art experiments have demonstrated the capability to prepare GKP states of squeezing between $5.5$dB and $9.5$dB \cite{Campagne2019,Fluhmann2018,Fluhmann2019,Fluhmann2019b}, approaching the established squeezing threshold values $s_{\textrm{gkp}}^{\star} \ge 11.2$dB.

In circuit QED systems, beam-splitter interactions between two high-Q cavity modes have been implemented experimentally with $\kappa/ g \sim 10^{-2}$, where $g$ is the relevant coupling strength and $\kappa$ is the photon loss rate \cite{Gao2018}. While the same scheme (based on four-wave mixing processes) may be adapted to realize the SUM or the inverse-SUM gates between two high-Q cavity modes \cite{Zhang2019}, this scheme will induce non-negligible Kerr nonlinearities and thus may not be compatible with the GKP qubits which should be operated in the regime where Kerr nonlinearities are negligible \cite{Campagne2019}. On the other hand, by using three-wave mixing elements \cite{Frattini2017}, it would be possible to implement the SUM or the inverse-SUM gates between two high-Q cavity modes in a way that is not significantly limited by Kerr nonlinearities.

Let us now compare the performance of the surface-GKP code with the usual rotated surface code implemented by bare qubits such as transmon qubits. Assuming a full circuit-level depolarizing noise (both for single- and two-qubit gates), it was numerically demonstrated that fault-tolerant quantum error correction is possible with the rotated surface code if the physical error rate is below the threshold $p^{\star} = 1.2\%$ \cite{Wang2011}. Note that such a high threshold value was obtained by introducing 3D space-time correlated edges (see Figs. 3 and 4 in Ref. \cite{Wang2011}) and fully optimizing the renormalized edge weights based on the noise parameters.  

Our circuit-level noise model (in terms of shift errors) is quite different from the depolarizing noise model considered in typical qubit-based fault-tolerant error correction schemes. Moreover, we also introduce non-Gaussian resources, i.e., GKP states in our scheme. Therefore, our results cannot be directly compared with the results in Ref. \cite{Wang2011}. We nevertheless point out that we obtain comparable threshold values $(\kappa/ g)^{\star} = 0.81\%$ (\ref{case:2}) and $(\kappa/ g)^{\star} = 0.69\%$ (\ref{case:3}) where $\kappa$ is the photon loss rate and $g$ is the coupling strength of the two-mode gates. We stress that we do not introduce 3D space-time correlated edges and provide a simple method for computing the renormalized edge weights. In particular, 3D space-time correlated edges are not necessary in our case with the surface-GKP code. This is because any shift errors that are correlated due to two-mode gates will not cause any Pauli errors to GKP qubits nor trigger syndrome GKP qubits incorrectly, as long as the size of the correlated shifts is smaller than $\sqrt{\pi}/2$, which is the case below the fault-tolerance thresholds computed above.

We also point out that in general, topological codes without leakage reduction units \cite{Aliferis2007} are not robust against leakage errors that occur when a bare qubit state is excited and falls out of its desired two-level subspace \cite{Aliferis2007,Fowler2013,Suchara2015,Brown2019}. In the case of the surface-GKP code, leakage errors do occur as well because each bosonic mode may not be in the desired two-level GKP code subspace. However, the surface-GKP code is inherently resilient to such leakage errors (and thus does not require leakage reduction units) since GKP-stabilizer measurements will detect and correct such events. Indeed, in our simulation of the surface-GKP code, leakage errors continuously occur due to shift errors, but the established fault-tolerance thresholds are nevertheless still favorable since GKP-stabilizer measurements prevent the leakage errors from propagating further.

\begin{figure}[t!]
\centering
\includegraphics[width=3.3in]{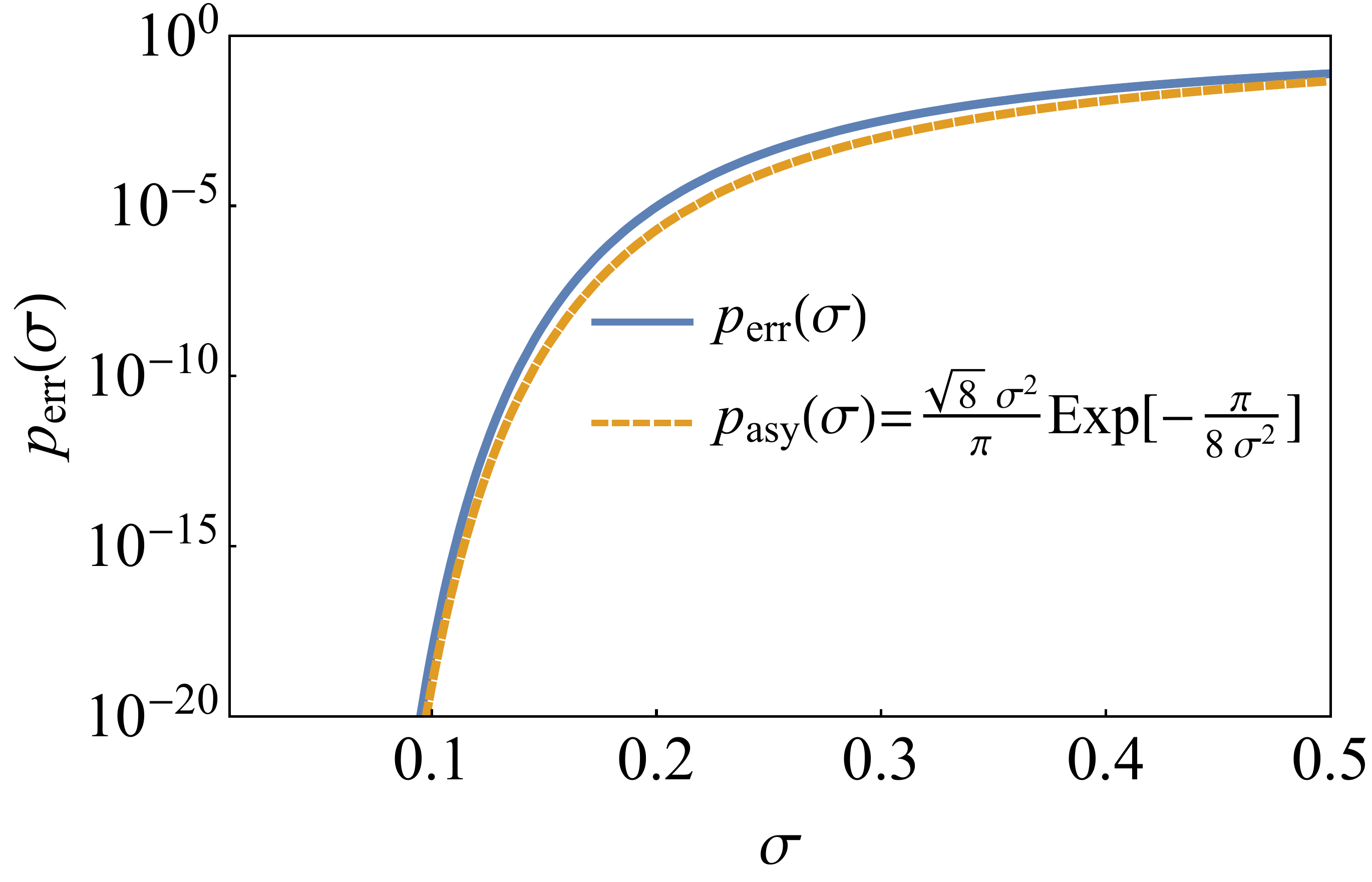}
\caption{Visualization of the function $p_{\textrm{err}}(\sigma)$ (blue). The asymptotic expression $p_{\textrm{asy}}(\sigma) = (\sqrt{8}\sigma^{2} / \pi ) \exp[ -\pi/ (8\sigma^{2} )  ]$ is represented by the yellow dashed line. $p_{\textrm{err}}(\sigma)$ and $p_{\textrm{asy}}(\sigma)$ agree well with each other in the $\sigma\ll 1$ limit.   }
\label{fig:perr}
\end{figure}

We lastly remark that the logical $X$ or $Z$ error rates in \cref{fig:main results} decrease very rapidly as $\sigma_{\textrm{gkp}}$ and $\sigma$ approach zero in the case of the surface-GKP code. This is again because the GKP code can correct any shift errors of size less than $\sqrt{\pi}/2$ and therefore the probability that a Pauli error occurs in a GKP qubit (at the end of GKP-stabilizer measurements) becomes exponentially small as $\sigma_{\textrm{gkp}}$ and $\sigma$ approach zero. More precisely, at the end of each GKP-stabilizer measurement, a bulk data GKP qubit undergoes a Pauli $X$ or $Z$ error with probability 
\begin{align}
p_{\textrm{err}}\Big{(} \sqrt{5\sigma_{\textrm{gkp}}^{2} + \frac{59}{3}\sigma^{2} } \Big{)}, 
\end{align}
where $p_{\textrm{err}}(\sigma)$ is defined in \cref{eq:definition of perr}. Here, the variance $5\sigma_{\textrm{gkp}}^{2} +  (59/3)\sigma^{2}$ was carefully determined by thoroughly keeping track of how circuit-level noise propagates during stabilizer measurements (see also \cref{section:Methods}). As can be seen from \cref{fig:perr}, $p_{\textrm{err}}(\sigma)$ agrees well with the asymptotic expression $p_{\textrm{asy}}(\sigma) = (\sqrt{8}\sigma^{2} / \pi ) \exp[ -\pi/ (8\sigma^{2} )  ]$ in the $\sigma\ll 1$ limit. Thus, $p_{\textrm{err}}(\sigma)$ decreases exponentially as $\sigma$ goes to zero. 

Similarly, the probability that a bulk surface code stabilizer measurement yields an incorrect measurement outcome is given by
\begin{align}
p_{\textrm{err}}\Big{(} \sqrt{7\sigma_{\textrm{gkp}}^{2} + \frac{116}{3}\sigma^{2} } \Big{)}, 
\end{align}
and decays exponentially as $\sigma_{\textrm{gkp}}$ and $\sigma$ approach zero. Therefore, if the circuit-level noise of the physical bosonic modes is very small to begin with, GKP codes will locally provide a significant noise reduction. In this case, the overall resource overhead associated with the next level of global encoding will be modest since a small-distance surface code would suffice. Therefore in this regime, the surface-GKP code may be able to achieve the same target logical error rate in a more hardware-efficient way than the usual surface code. However, since this regime requires high quality GKP states, the additional resource overhead associated with the preparation of such high quality GKP states should also be taken into account for a comprehensive resource estimate. We leave such an analysis to future work.      

\section{Discussion and outlook}
\label{section:Discussion}

\begin{table*}[t!]
  \centering
  \def\arraystretch{2}
  \begin{tabular}{ V{3} c V{1.5} c V{1.5}  c V{1.5} c V{1.5} c V{1.5} c V{3} }
   \hlineB{3}  
    \ref{case:1} ($\sigma=0$) & Method & \,\,\, $\sigma_{\textrm{gkp}}^{\star}$ \,\,\, & \,\,\, $s_{\textrm{gkp}}^{\star}$ \,\,\, &  Post-selection? & Success probability \\  \hlineB{3} 
    Ref. \cite{Menicucci2014} & Concatenated codes (MB) & $0.067$ & $20.5$dB &   NO & $1$ \\ \hlineB{1.5}  
    Ref. \cite{Fukui2018a} & 3D cluster state (MB) & $0.228$ & $9.8$dB &  YES & decreases exponentially with $d^{3}$ \\ \hlineB{1.5}  
    Ref. \cite{Fukui2019} & 3D cluster state (MB) & $0.273$ & $8.3$dB &  YES & decreases exponentially with $d^{3}$  \\ \hlineB{1.5}  
    Refs. \cite{Wang2017,Vuillot2019} & Toric-GKP code (GB) & N/A  & N/A &  NO  & 1 \\ \hlineB{1.5}  
    Our work & Surface-GKP code (GB) & $0.194$ & $11.2$dB &  NO & 1  \\ \hlineB{3}  
  \end{tabular}
   \caption{Threshold values for the squeezing of GKP states for fault-tolerant quantum error correction. Here, we compare the established threshold values obtained by assuming that 
   GKP states are the only noisy components in the error correction circuit (i.e., \ref{case:1}). MB stands for measurement-based and GB stands for gate-based. $d$ is the distance of the code. For Refs. \cite{Wang2017,Vuillot2019}, $\sigma_{\textrm{gkp}}^{\star}$ and $s_{\textrm{gkp}}^{\star}$ are not available because they assumed phenomenological noise models that do not take into account the propagation of shift errors through the entire error correction circuit. That is, the threshold values established in Ref. \cite{Vuillot2019} using the toric-GKP code (i.e., $\sigma_{0}^{\star} = 0.235$ and $\sigma_{0}^{\star} = 0.243$ see Fig. 12 therein) do not accurately quantify the tolerable noise in the ancilla GKP states. Instead, $\sigma_{0}^{\star}$ can only be taken as a rough upper bound on $\sigma_{\textrm{gkp}}^{\star}$ (see the main text for more details).  }
   \label{table:comparison with other works}
\end{table*}

Here, we compare the results obtained in this paper with previous works in \cite{Menicucci2014,Wang2017,Fukui2018a,Fukui2019,Vuillot2019}. Firstly, Refs. \cite{Wang2017,Vuillot2019} considered the toric-GKP code and computed fault-tolerance thresholds for both code capacity and phenomenological noise models. In particular, the phenomenological noise models used in these works describe faulty syndrome extraction procedures (due to finitely-squeezed ancilla GKP states) in a way that does not take into account the propagation of the relevant shift errors. More specifically, in Figs. 1, 2, and 7 in Ref. \cite{Vuillot2019}, shift errors are manually added in the beginning of each stabilizer measurement measurement and right before each homodyne measurement. Therefore, this phenomenological noise model can be understood as a model for homodyne detection inefficiencies while assuming ideal ancilla GKP states. In other words, the fault-tolerance threshold values established in Ref. \cite{Vuillot2019} (i.e., $\sigma_{0}^{\star} = 0.235$ and $\sigma_{0}^{\star} = 0.243$; see Fig. 12 therein) do not accurately represent the tolerable noise in the ancilla GKP states since the noise propagation was not thoroughly taken into account. Thus, these threshold values can only be taken as a rough upper bound on $\sigma_{\textrm{gkp}}^{\star}$ and cannot be directly compared with the threshold values obtained in our work. Note also that the threshold values in Ref. \cite{Vuillot2019} were computed for the toric code which has a different threshold compared to the rotated surface code \cite{Fowler2012}. 


On the other hand, in our work we assume that every GKP state supplied to the error correction chain has a finite squeezing and we comprehensively take into account the propagation of such shift errors through the entire error correction circuit. By doing so, we accurately estimate the tolerable noise in the finitely-squeezed ancilla GKP states by computing $\sigma_{\textrm{gkp}}^{\star}$. Related, we stress that when the noise propagation is taken into account, detailed scheduling and design of the syndrome extraction circuits become very crucial and we carefully designed the circuits in a way that mitigates the adverse effects of the noise propagation (see \cref{fig:Surface code error propagation}).    

Moreover, we also consider photon loss and heating errors occurring continuously during the implementation of the SUM and inverse-SUM gates.
Thus, we establish fault-tolerance thresholds for the strength of the two-mode coupling relative to the photon loss rate and demonstrate that fault-tolerant quantum error correction with the surface-GKP code is possible in more general scenarios. We also remark that Ref. \cite{Vuillot2019} used a minimum-energy decoder based on statistical-mechanical methods in the noisy regime whereas we provide a simple method for computing renormalized edge weights to be used in a MWPM decoder.


Secondly, Refs. \cite{Menicucci2014,Fukui2018a,Fukui2019} considered measurement-based quantum computing with GKP qubits and did establish fault-tolerance thresholds for the squeezing of the GKP states. Assuming that GKP states are the only noisy components (i.e., \ref{case:1}), Ref. \cite{Menicucci2014} found the squeezing threshold value $s_{\textrm{gkp}}^{\star} = 20.5$dB, and Refs. \cite{Fukui2018a} and \cite{Fukui2019} later brought the value down to $s_{\textrm{gkp}}^{\star} = 9.8$dB and $s_{\textrm{gkp}}^{\star} = 8.3$dB, respectively. Notably, the squeezing thresholds found in Refs. \cite{Fukui2018a,Fukui2019} are more favorable than the squeezing threshold found in our work, i.e., $s_{\textrm{gkp}}^{\star} = 11.2$dB (see \cref{fig:main results} (a)). In this regard, we remark that the favorable threshold values obtained in Refs. \cite{Fukui2018a,Fukui2019} rely on the use of post-selection. That is, each GKP measurement succeeds with probability strictly less than unity and thus the overall success probability would decrease exponentially as the system size $d$ increases. On the other hand, we do not discard any measurement outcomes and thus our scheme succeeds with unit probability for any distance $d$. Therefore, our scheme with the surface-GKP code deterministically suppresses errors exponentially with the code distance as long as $\sigma_{\textrm{gkp}}$ and $\sigma$ are below the threshold values. The differences between our work and the previous works are summarized in \cref{table:comparison with other works}.

Let us now consider the number of bosonic modes needed to implement the distance-$d$ surface-GKP code: Recall \cref{fig:Surface-GKP codes} and note that we use $d^{2}$ data modes (white circles in \cref{fig:Surface-GKP codes}), $d^{2}$ ancilla modes (grey circles in \cref{fig:Surface-GKP codes}), and $d^{2}-1$ syndrome modes (green and orange circles in \cref{fig:Surface-GKP codes}). Although we introduced the $d^{2}$ ancilla modes to describe our scheme in a simpler way, the $d^{2}$ ancilla modes can in fact be replaced by the $d^{2}-1$ syndrome qubits plus one more additional mode. Thus, we only need a total of $2d^{2}$ modes and geometrically local two-mode couplings to implement the distance-$d$ surface-GKP code. For example, $18$ modes would suffice to realize the smallest non-trivial case with $d=3$.

We finally emphasize that we modeled noisy GKP states by applying an incoherent random displacement error $\mathcal{N}[\sigma_{\textrm{gkp}}]$ to the ideal GKP states, similarly as in Refs. \cite{Menicucci2014,Fukui2018a,Fukui2019}. While we use this noise model for theoretical convenience and justify it by using a twirling argument (see \cref{appendix:Supplementary material for noisy GKP states}), similar to the justification of a depolarizing error model in qubit-based QEC, we remark that it is not practical to use the twirling operation in realistic situations. This is because the twirling operation increases the average photon number of the GKP states, whereas in practice it is desirable to keep the photon number bounded below a certain cutoff. Therefore, an interesting direction for future work would be to see if one can implement the stabilizer measurements in \cref{fig:GKP qubit fundamentals,fig:Surface-GKP codes,fig:Surface code stabilizer measurement} in a manner that prevents the average photon number from diverging as we repeat the stabilizer measurements. It will be especially crucial to keep the average photon number under control when each bosonic mode suffers from dephasing errors and/or undesired nonlinear interactions such as Kerr nonlinearities.

Related, we remark that in the recent experimental realization of the GKP code in a circuit QED system, an envelope-trimming technique was used to constrain the average photon number of the system \cite{Campagne2019}. Whether a similar technique can be incorporated in a large-scale surface-GKP code architecture would be an interesting future research direction.

To summarize, we have thoroughly investigated the performance of the surface-GKP code assuming a detailed circuit-level noise model. By simulating the full noisy error correction protocol and using a minimum-weight perfect matching decoding on a 3D space-time graph (with a simple method for computing renormalized edge weights), we numerically demonstrated that fault-tolerant quantum error correction is possible with the surface-GKP code if the squeezing of the GKP states and the circuit noise are below certain fault-tolerance thresholds. Since our scheme does not require any post-selection and thus succeeds with unit probability, our scheme is clearly scalable. We also described our methods in great detail such that our results can easily be reproduced.

\section*{Acknowledgments}

We thank Andrew Cross, Christophe Vuillot, Barbara Terhal, Alec Eickbusch, Steven Touzard, and Philippe Campagne-Ibarcq for helpful discussions and for providing comments on the manuscript. K.N. is grateful for the hospitality of the IBM T.J. Watson research center where this work was conceived and completed. 


\appendix

\section{Supplementary material for the noise model}
\label{appendix:Supplementary material for noisy GKP states}

To derive \cref{eq:finite GKP expansion in displacements}, we used the following identity. 
\begin{align}
&\textrm{Tr}\big{[} \exp[-\Delta\hat{n}] \hat{D}^{\dagger}(\alpha)\big{]} 
\nonumber\\
&= \sum_{n=0}^{\infty}e^{-\Delta n} \langle n| \hat{D}^{\dagger}(\alpha)|n\rangle 
\nonumber\\
&= \exp\Big{[} -\frac{|\alpha|^{2}}{2} \Big{]} \sum_{n=0}^{\infty} e^{-\Delta n} L_{n}(|\alpha|^{2}) 
\nonumber\\
&= \exp\Big{[} -\frac{|\alpha|^{2}}{2} \Big{]} \frac{1}{1-e^{-\Delta}} \exp\Big{[}-\frac{ e^{-\Delta} }{ 1-e^{-\Delta} } |\alpha|^{2} \Big{]} 
\nonumber\\
&= \frac{1}{1-e^{-\Delta}} \exp\Big{[} -\frac{ 1+e^{-\Delta} }{2( 1-e^{-\Delta}) } |\alpha|^{2} \Big{]}, \label{appeq:laguerre identity}
\end{align}
where $L_{n}(x)$ is the Laguerre polynomial. Further, going from the third to fourth line, we used the generating function for the Laguerre polynomials which satisfies $\sum_{n=0}^{\infty} t^{n}L_{n}(x) = \frac{1}{1-t} e^{ -tx/(1-t)}$.

\begin{widetext}
Now, we explain how one can transform the noisy GKP state corrupted by coherent superpositions of displacement errors (see \cref{eq:finite GKP expansion in displacements}) into the a noisy GKP state corrupted by an incoherent mixture of displacement errors (see \cref{eq:noisy GKP state incoherent displacement error}). To do so, we apply random shifts of integer multiples $2\sqrt{\pi}$ in both the position and the momentum directions to the noisy GKP state $|\psi^{\Delta}_{\textrm{gkp}}\rangle\propto \exp[-\Delta \hat{n}] |\psi_{\textrm{gkp}}\rangle$. Then, $|\psi^{\Delta}_{\textrm{gkp}}\rangle$ is transformed into 
\begin{align}
\hat{\psi}_{\textrm{gkp}}^{\Delta}  &\propto \sum_{n_{1},n_{2}\in \mathbb{Z}} (\hat{S}_{q})^{n_{1}} (\hat{S}_{p})^{n_{2}} |\psi^{\Delta}_{\textrm{gkp}}\rangle\langle \psi_{\textrm{gkp}}^{\Delta}|  (\hat{S}_{p}^{\dagger})^{n_{2}} (\hat{S}_{q}^{\dagger})^{n_{1}}
\nonumber\\
&\propto \sum_{n_{1},n_{2}\in \mathbb{Z}} \int d^{2}\alpha d^{2}\beta \exp\Big{[} -\frac{ |\alpha|^{2} + |\beta|^{2} }{ 2\sigma_{\textrm{gkp}}^{2} } \Big{]}  (\hat{S}_{q})^{n_{1}} (\hat{S}_{p})^{n_{2}} \hat{D}(\alpha) |\psi_{\textrm{gkp}}\rangle\langle \psi_{\textrm{gkp}}| \hat{D}^{\dagger}(\beta)  (\hat{S}_{p}^{\dagger})^{n_{2}} (\hat{S}_{q}^{\dagger})^{n_{1}}
\nonumber\\
&\propto \sum_{n_{1},n_{2}\in \mathbb{Z}} \int d^{2}\alpha d^{2}\beta \exp\Big{[} -\frac{ |\alpha|^{2} + |\beta|^{2} }{ 2\sigma_{\textrm{gkp}}^{2} } \Big{]} \exp[ i\sqrt{2\pi}(\alpha_{R}-\beta_{R})n_{1} -i\sqrt{2\pi}(\alpha_{I}-\beta_{I})n_{2} ]
\nonumber\\
&\qquad\qquad\qquad\qquad\qquad\qquad\qquad\qquad\qquad \times \hat{D}(\alpha) (\hat{S}_{q})^{n_{1}} (\hat{S}_{p})^{n_{2}}  |\psi_{\textrm{gkp}}\rangle\langle \psi_{\textrm{gkp}}|  (\hat{S}_{p}^{\dagger})^{n_{2}} (\hat{S}_{q}^{\dagger})^{n_{1}}\hat{D}^{\dagger}(\beta)  
\nonumber\\
&\propto \sum_{n_{1},n_{2}\in \mathbb{Z}} \int d^{2}\alpha d^{2}\beta \exp\Big{[} -\frac{ |\alpha|^{2} + |\beta|^{2} }{ 2\sigma_{\textrm{gkp}}^{2} } \Big{]} \exp[ i\sqrt{2\pi}(\alpha_{R}-\beta_{R})n_{1} -i\sqrt{2\pi}(\alpha_{I}-\beta_{I})n_{2} ]  \hat{D}(\alpha) |\psi_{\textrm{gkp}}\rangle\langle \psi_{\textrm{gkp}}|   \hat{D}^{\dagger}(\beta)  , \label{appeq:GKP twirling intermediate}
\end{align}
where we used the identity $\hat{D}(\alpha)\hat{D}(\beta) = \hat{D}(\beta)\hat{D}(\alpha) e^{\alpha\beta^{*}-\alpha^{*}\beta}$ and the fact that GKP states are stabilized by the GKP stabilizers $\hat{S}_{q} = \hat{D}(i\sqrt{2\pi})$ and $\hat{S}_{p} = \hat{D}(\sqrt{2\pi})$, i.e., $\hat{S}_{q}|\psi_{\textrm{gkp}}\rangle = \hat{S}_{p}|\psi_{\textrm{gkp}}\rangle =|\psi_{\textrm{gkp}}\rangle$. Using the Poisson summation formula, $\sum_{n\in\mathbb{Z}} e^{i a n}  = 2\pi \sum_{k\in\mathbb{Z}} \delta(a-2\pi k)$ we can further simplify \cref{appeq:GKP twirling intermediate} as 
\begin{align}
\hat{\psi}_{\textrm{gkp}}^{\Delta}  &\propto \sum_{k_{1},k_{2}\in \mathbb{Z}} \int d^{2}\alpha d^{2}\beta \exp\Big{[} -\frac{ |\alpha|^{2} + |\beta|^{2} }{ 2\sigma_{\textrm{gkp}}^{2} } \Big{]} \delta(\alpha_{R}-\beta_{R} - \sqrt{2\pi}k_{1} )\delta(\alpha_{I}-\beta_{I} - \sqrt{2\pi}k_{2} )  \hat{D}(\alpha) |\psi_{\textrm{gkp}}\rangle\langle \psi_{\textrm{gkp}}|   \hat{D}^{\dagger}(\beta) 
\nonumber\\
&=  \sum_{k_{1},k_{2}\in \mathbb{Z}} \int d^{2}\alpha  \exp\Big{[} -\frac{ |\alpha|^{2} + |\alpha -\sqrt{2\pi} (k_{1}+ ik_{2}) |^{2} }{ 2\sigma_{\textrm{gkp}}^{2} } \Big{]}  \hat{D}(\alpha) |\psi_{\textrm{gkp}}\rangle\langle \psi_{\textrm{gkp}}|   \hat{D}^{\dagger}(\alpha -\sqrt{2\pi} (k_{1}+ik_{2}) ) 
\nonumber\\
&=  \sum_{k_{1},k_{2}\in \mathbb{Z}}   \exp\Big{[} -\frac{ \pi |k_{1}+ ik_{2}|^{2}      }{ 2\sigma_{\textrm{gkp}}^{2} } \Big{]}  \int d^{2}\alpha \exp\Big{[} -\frac{ |\alpha - \sqrt{\frac{\pi}{2}} (k_{1} + ik_{2}) |^{2}  }{\sigma_{\textrm{gkp}}^{2} } \Big{]}  \hat{D}(\alpha) |\psi_{\textrm{gkp}}\rangle\langle \psi_{\textrm{gkp}}|   \hat{D}^{\dagger}(\alpha -\sqrt{2\pi} (k_{1}+ik_{2}) ) . 
\end{align}
Lastly, if $\sigma_{\textrm{gkp}} \ll \sqrt{\pi}$ (which is the case below the fault-tolerance threshold $\sigma_{\textrm{gkp}}^{\star} \le 0.194$), we can neglect all the $(k_{1},k_{2}) \neq (0,0)$ terms due to the exponentially decaying prefactor $ \exp[ -\frac{ \pi |k_{1}+ ik_{2}|^{2}      }{ 2\sigma_{\textrm{gkp}}^{2} } ] $ and get the noise model in \cref{eq:noisy GKP state incoherent displacement error}:
\begin{align}
\hat{\psi}_{\textrm{gkp}}^{\Delta}  &\propto   \int \frac{d^{2}\alpha }{ \pi\sigma_{\textrm{gkp}}^{2} } \exp\Big{[} -\frac{ |\alpha  |^{2}  }{\sigma_{\textrm{gkp}}^{2} } \Big{]}  \hat{D}(\alpha) |\psi_{\textrm{gkp}}\rangle\langle \psi_{\textrm{gkp}}|   \hat{D}^{\dagger}(\alpha  )  = \mathcal{N}[\sigma_{\textrm{gkp}}]( |\psi_{\textrm{gkp}}\rangle\langle \psi_{\textrm{gkp}}|   ) . 
\end{align}

\end{widetext}

Let us now derive the gate error model given in \cref{eq:noise covariance matrix SUM or inverse-SUM}. Recall that $\mathcal{L}'_{\pm}$ is given by 
\begin{align}
\mathcal{L}'_{\pm} = \mathcal{V}_{\pm} + \mathcal{L}_{\textrm{err}},  
\end{align} 
where $\mathcal{V}_{\pm}$ and $\mathcal{L}_{\textrm{err}}$ are defined as 
\begin{align}
\mathcal{V}_{\pm}(\hat{\rho}) &\equiv  \mp ig [ \hat{q}_{1}\hat{p}_{2} , \hat{\rho} ], 
\nonumber\\
\mathcal{L}_{\textrm{err}}(\hat{\rho}) &\equiv \kappa \sum_{k=1}^{2} \big{(} \mathcal{D}[\hat{a}_{k}] + \mathcal{D}[\hat{a}^{\dagger}_{k}] \big{)}\hat{\rho}. 
\end{align}
The noisy SUM or the inverse-SUM gates is then given by $\exp[\mathcal{L}'_{\pm} \Delta t]$ with $\Delta t = 1/g$. Note that Trotter's formula \cite{Trotter1959} yields 
\begin{align}
\exp[\mathcal{L}'_{\pm} \Delta t] &= \lim_{N\rightarrow \infty} \Big{[}  \exp\Big{[} \mathcal{V}_{\pm} \frac{\Delta t}{N} \Big{]} \exp\Big{[} \mathcal{L}_{\textrm{err}} \frac{\Delta t}{N} \Big{]}  \Big{]}^{N} . 
\end{align} 
Note that both $\exp [  \mathcal{V}_{\pm} \Delta t/N ] $ and $\exp[ \mathcal{L}_{\textrm{err}} \Delta t/ N]$ are Gaussian channels with the characterization matrices 
\begin{align}
\boldsymbol{T}_{\pm} &= \begin{bmatrix}
1&0&0&0\\
0&1&0&\mp 1/N\\
\pm 1/N&0&1&0\\
0&0&0&1
\end{bmatrix}, \quad \boldsymbol{N}_{\pm} =0, 
\nonumber\\
\boldsymbol{T}_{\textrm{err}} &= \begin{bmatrix}
1&0&0&0\\
0&1&0&0\\
0&0&1&0\\
0&0&0&1
\end{bmatrix}, \quad \boldsymbol{N}_{\textrm{err}} = \frac{\kappa\Delta t}{N}\begin{bmatrix}
1&0&0&0\\
0&1&0&0\\
0&0&1&0\\
0&0&0&1
\end{bmatrix}, 
\end{align}
respectively (see, for example, Ref. \cite{Weedbrook2012} for the definition of Gaussian channels and their characterization matrices). Thus, the quadrature operator $\boldsymbol{\hat{x}}  = ( \hat{q}_{1}, \hat{p}_{1} , \hat{q}_{2}, \hat{p}_{2} )^{T}$ is transformed via the noisy SUM or the inverse-SUM gate as 
\begin{align}
\boldsymbol{\hat{x}} &\rightarrow (\boldsymbol{T}_{\pm})^{N}\boldsymbol{\hat{x}} = \begin{bmatrix}
1&0&0&0\\
0&1&0&\mp 1\\
\pm 1&0&1&0\\
0&0&0&1
\end{bmatrix} \boldsymbol{\hat{x}} = \begin{bmatrix}
\hat{q}_{1}\\
\hat{p}_{1} \mp \hat{p}_{2}\\
\hat{q}_{2} \pm \hat{q}_{1}\\
\hat{p}_{2}
\end{bmatrix}  , 
\end{align}
as desired. Also, the covariance matrix $\boldsymbol{V}$ is transformed as 
\begin{align}
\nonumber\\
\boldsymbol{V} &\rightarrow (\boldsymbol{T}_{\pm})^{N} \boldsymbol{V} ( (\boldsymbol{T}_{\pm})^{N})^{T}  + \sum_{k=1}^{N} (\boldsymbol{T}_{\pm})^{k}\boldsymbol{N}_{\textrm{err}}(\boldsymbol{T}_{\pm}^{T})^{k}
\nonumber\\
&=  (\boldsymbol{T}_{\pm})^{N} \boldsymbol{V} ( (\boldsymbol{T}_{\pm})^{N})^{T}  
\nonumber\\
&\quad+ \sum_{k=1}^{N} \frac{\kappa \Delta t}{N} \begin{bmatrix}
1&0&\pm \frac{k}{N}&0\\
0&1+(\frac{k}{N})^{2}&0&\mp \frac{k}{N}\\
\pm \frac{k}{N}&0&1+(\frac{k}{N})^{2}&0\\
0&\mp \frac{k}{N} &0&1
\end{bmatrix}
\nonumber\\
&=(\boldsymbol{T}_{\pm})^{N} \boldsymbol{V} ( (\boldsymbol{T}_{\pm})^{N})^{T}  
\nonumber\\
&\quad+ \kappa \Delta t \begin{bmatrix}
1&0&\pm 1/2&0\\
0&4/3&0&\mp 1/2\\
\pm 1/2 &0&4/3&0\\
0&\mp 1/2 &0&1
\end{bmatrix}. 
\end{align}
Therefore, the noisy SUM or the inverse-SUM gate can be understood as the ideal SUM or the inverse-SUM gate followed by a correlated Gaussian random displacement error with the noise covariance matrices $\boldsymbol{N}_{q}^{\pm}$ and $\boldsymbol{N}_{p}^{\pm}$ as given in \cref{eq:noise covariance matrix SUM or inverse-SUM}.

\section{Simulation details} 
\label{section:Methods}

Here, we describe in detail how we simulate the syndrome extraction protocol for the surface-GKP code and how we decode the obtained syndrome measurement outcome. 

\begin{figure*}
\centering
\includegraphics[width=5.8in]{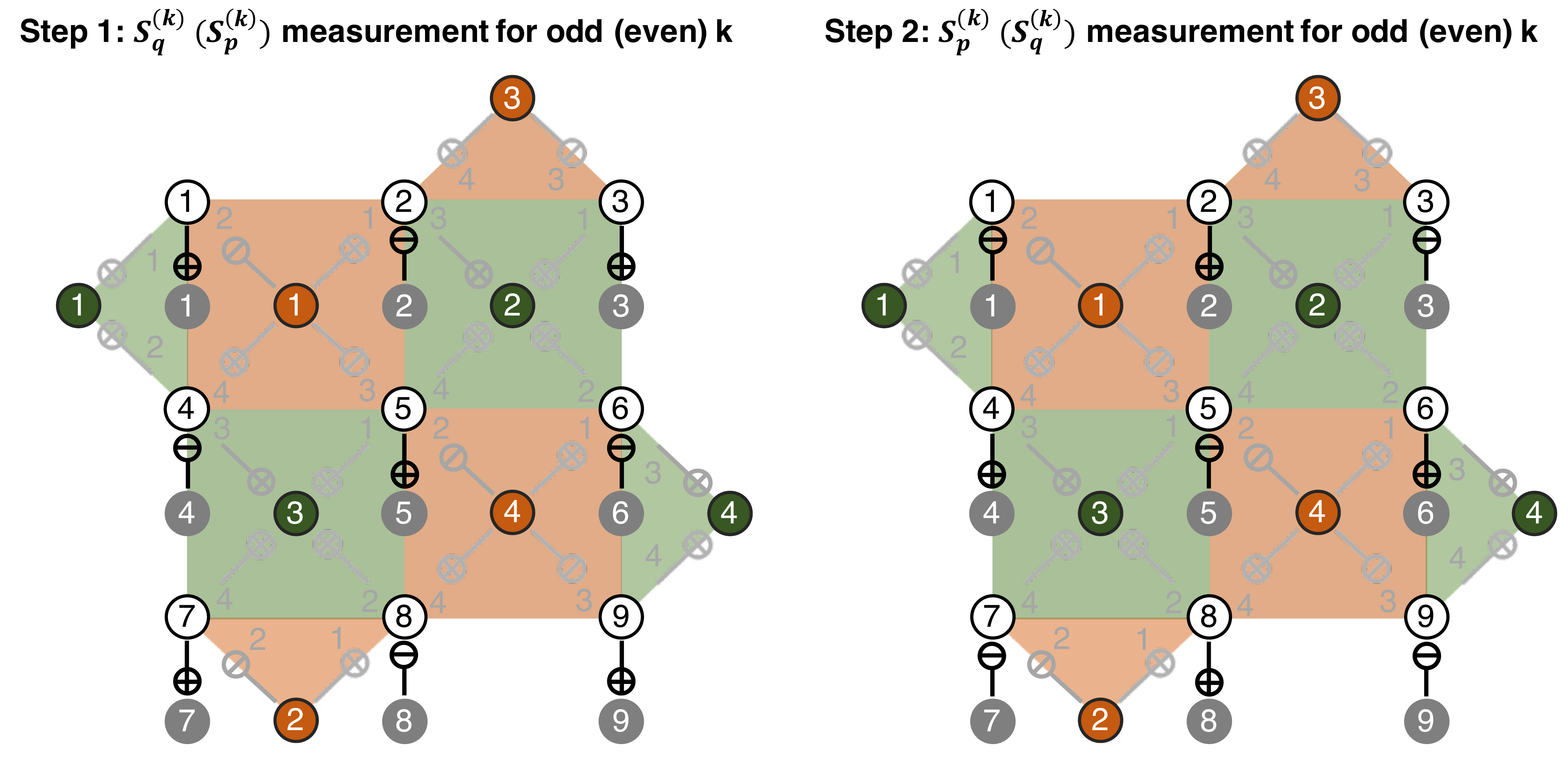}
\caption{Measurement of the GKP stabilizers for $d=3$. See also \cref{fig:GKP qubit fundamentals} (b) for the graphical notation.  }
\label{fig:GKP stabilizer measurement}
\end{figure*}

\subsection*{GKP-stabilizer measurements}
\label{subsection:GKP stabilizer measurement}

Consider the distance-$d$ surface-GKP code consisting of $d^{2}$ data GKP qubits. Each data GKP qubit is stabilized by the two GKP stabilizers $\hat{S}_{q}^{(k)} = \exp[i2\sqrt{\pi} \hat{q}_{k} ]$ and $\hat{S}_{p}^{(k)} =  \exp[- i2\sqrt{\pi} \hat{p}_{k} ]$ where $k\in \lbrace 1, \cdots, d^{2} \rbrace$. In the first step of GKP-stabilizer measurements (left in \cref{fig:GKP stabilizer measurement}), $\hat{S}_{q}^{(k)}$ ($\hat{S}_{p}$) stabilizers are measured for odd (even) $k$. In the second step (right in \cref{fig:GKP stabilizer measurement}), on the other hand, $\hat{S}_{p}^{(k)}$ ($\hat{S}_{q}^{(k)}$) stabilizers are measured for odd (even) $k$. Note that we alternate between $\hat{S}_{q}$ and $\hat{S}_{p}$ measurements in a checkerboard pattern in order to balance the position and momentum quadrature noise.

Let $\xi_{q}^{D}$ and $\xi_{p}^{D}$ ($\xi_{q}^{A}$ and $\xi_{p}^{A}$) be the data (ancilla) position and momentum quadrature noise, where 
\begin{align}
\xi_{q}^{D} &= (\xi_{q}^{(D1)},\cdots, \xi_{q}^{(Dd^{2})}),
\nonumber\\
\xi_{p}^{D} &= (\xi_{p}^{(D1)},\cdots, \xi_{p}^{(Dd^{2})}), 
\nonumber\\
\xi_{q}^{A} &= (\xi_{q}^{(A1)},\cdots, \xi_{q}^{(Ad^{2})}),
\nonumber\\
\xi_{p}^{A} &= (\xi_{p}^{(A1)},\cdots, \xi_{p}^{(Ad^{2})}). 
\end{align}
In Step 1, we add random shift errors occurring during the GKP state preparation as follows: 
\begin{align}
\xi_{q}^{(Dk)} &\leftarrow \xi_{q}^{(Dk)}  + \textrm{randG}(\sigma^{2}), 
\nonumber\\
\xi_{p}^{(Dk)} &\leftarrow \xi_{p}^{(Dk)}  + \textrm{randG}(\sigma^{2}),
\nonumber\\
\xi_{q}^{(Ak)} &\leftarrow  \textrm{randG}(\sigma_{\textrm{gkp}}^{2} ), 
\nonumber\\
\xi_{p}^{(Ak)} &\leftarrow  \textrm{randG}(\sigma_{\textrm{gkp}}^{2} ), \label{eq:noise during gkp preparation} 
\end{align}
for $k \in\lbrace 1,\cdots, d^{2} \rbrace$ where $\textrm{randG}(\boldsymbol{V})$ generates a random vector sampled from a multivariate Gaussian distribution $\mathcal{N}(0, \boldsymbol{V})$ with zero mean and the covariance matrix $\boldsymbol{V}$. Then, due to the SUM and the inverse-SUM gates, the quadrature noise vectors are updated as follows. 
\begin{align}
(\xi_{q}^{(Dk)} , \xi_{q}^{(Ak)} ) &\leftarrow  (\xi_{q}^{(Dk)} , \xi_{q}^{(Ak)} +\xi_{q}^{(Dk)} ) 
\nonumber\\
&\qquad + \textrm{randG}\Big{(} \sigma^{2}\begin{bmatrix}
1 & 1/2 \\
1/2 & 4/3
\end{bmatrix} \Big{)} , 
\nonumber\\
(\xi_{p}^{(Dk)} , \xi_{p}^{(Ak)} ) &\leftarrow  (\xi_{p}^{(Dk)} - \xi_{p}^{(Ak)} , \xi_{p}^{(Ak)} ) 
\nonumber\\
&\qquad + \textrm{randG}\Big{(} \sigma^{2}\begin{bmatrix}
4/3 & -1/2 \\
-1/2 & 1
\end{bmatrix} \Big{)} , \label{eq:SUM between data and ancilla}
\end{align} 
for odd $k$ ($\hat{S}_{q}^{(k)}$ stabilizer measurement) and 
\begin{align}
(\xi_{q}^{(Dk)} , \xi_{q}^{(Ak)} ) &\leftarrow  (\xi_{q}^{(Dk)} - \xi_{q}^{(Ak)} , \xi_{q}^{(Ak)} ) 
\nonumber\\
&\qquad + \textrm{randG}\Big{(} \sigma^{2}\begin{bmatrix}
4/3 & -1/2 \\
-1/2 & 1
\end{bmatrix} \Big{)} , 
\nonumber\\
(\xi_{p}^{(Dk)} , \xi_{p}^{(Ak)} ) &\leftarrow  (\xi_{p}^{(Dk)} , \xi_{p}^{(Ak)} + \xi_{p}^{(Dk)} ) 
\nonumber\\
&\qquad + \textrm{randG}\Big{(} \sigma^{2}\begin{bmatrix}
1 & 1/2 \\
1/2 & 4/3
\end{bmatrix} \Big{)} ,  \label{eq:inverse-SUM between data and ancilla}
\end{align} 
for even $k$ ($\hat{S}_{p}^{(k)}$ stabilizer measurement). Due to the noise before (or during) the homodyne measurement, the noise vectors are updated as 
 \begin{align}
\xi_{q}^{(Dk)} &\leftarrow \xi_{q}^{(Dk)}  + \textrm{randG}(\sigma^{2}), 
\nonumber\\
\xi_{p}^{(Dk)} &\leftarrow \xi_{p}^{(Dk)}  + \textrm{randG}(\sigma^{2}),
\nonumber\\
\xi_{q}^{(Ak)} &\leftarrow \xi_{q}^{(Ak)}  + \textrm{randG}(\sigma^{2} ), 
\nonumber\\
\xi_{p}^{(Ak)} &\leftarrow \xi_{p}^{(Ak)}  + \textrm{randG}(\sigma^{2} ), \label{eq:noise during homodyne measurement}
\end{align}
for all $k\in \lbrace 1,\cdots, d^{2} \rbrace$. Then, through the homodyne measurement and the error correction process, the data noise vectors are transformed as 
\begin{align}
\xi_{q}^{(Dk)} \leftarrow \xi_{q}^{(Dk)}  - R_{\sqrt{\pi}} \big{(} \xi_{q}^{(Ak)} \big{)} , \label{eq:position correction}
\\
\xi_{p}^{(Dk)} \leftarrow \xi_{p}^{(Dk)}  - R_{\sqrt{\pi}} \big{(} \xi_{p}^{(Ak)} \big{)} , \label{eq:momentum correction}
\end{align}
for odd and even $k$, respectively. $R_{s}(z)$ is defined as 
\begin{align}
R_{s}(z) \equiv z - s \Big{ \lfloor} \frac{z}{s}+ \frac{1}{2} \Big{\rfloor} . \label{eq:definition of the R function}
\end{align}

In Step 2, $\hat{S}_{p}^{(k)}$ ($\hat{S}_{q}^{(k)}$) stabilizers are measured for odd (even) $k$ instead of $\hat{S}_{q}^{(k)}$ ($\hat{S}_{p}^{(k)}$). Thus, the noise vectors are updated similarly as in \cref{eq:noise during gkp preparation,eq:SUM between data and ancilla,eq:inverse-SUM between data and ancilla,eq:noise during homodyne measurement,eq:position correction,eq:momentum correction}, except that \cref{eq:SUM between data and ancilla,eq:position correction} (\cref{eq:inverse-SUM between data and ancilla,eq:momentum correction}) are applied when $k$ is even (odd) instead of when $k$ is odd (even).

\subsection*{Surface code stabilizer measurements} 
\label{subsection:Surface code stabilizer measurement}

\begin{figure*}
\centering
\includegraphics[width=6.0in]{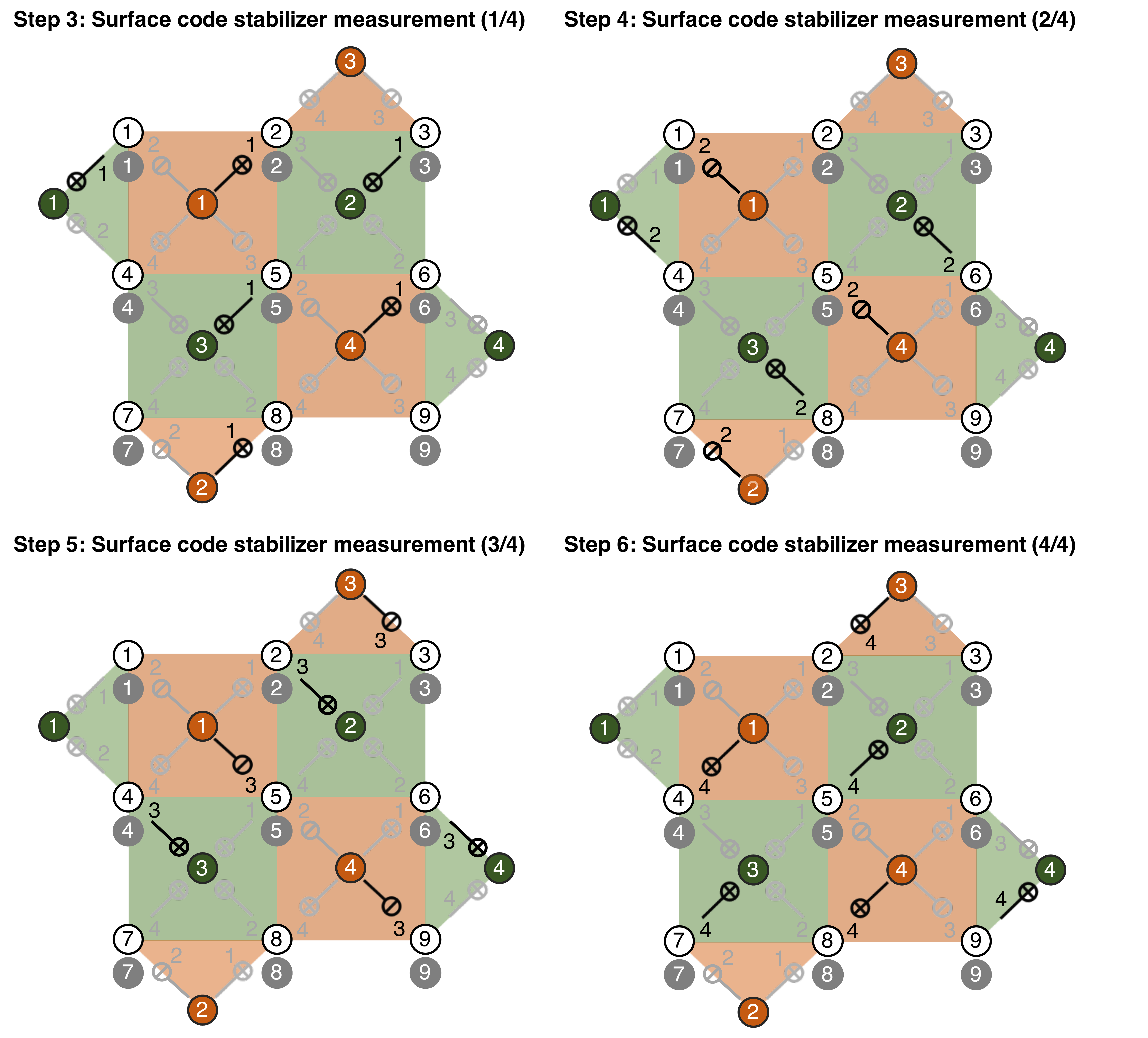}
\caption{Measurement of the surface code stabilizers for $d=3$.  }
\label{fig:surface code stabilizer measurement four time steps}
\end{figure*}

Recall that there are $d'\equiv (d^{2}-1)/2$ $Z$-type and $X$-type syndrome GKP qubits that are used to measure the surface code stabilizers. Let $\xi_{q}^{Z}$ and $\xi_{p}^{Z}$ ($\xi_{q}^{X}$ and $\xi_{p}^{X}$) be the position and momentum noise vectors of the $Z$-type ($X$-type) syndrome GKP qubits, where 
\begin{align}
\xi_{q}^{Z} &= (\xi_{q}^{(Z1)},\cdots, \xi_{q}^{(Zd')}),
\nonumber\\
\xi_{p}^{Z} &= (\xi_{p}^{(Z1)},\cdots, \xi_{p}^{(Zd')}), 
\nonumber\\
\xi_{q}^{X} &= (\xi_{q}^{(X1)},\cdots, \xi_{q}^{(Xd')}),
\nonumber\\
\xi_{p}^{X} &= (\xi_{p}^{(X1)},\cdots, \xi_{p}^{(Xd')}). 
\end{align}
Note that the SUM and the inverse-SUM gates for the syndrome extraction are executed in four time steps (see Steps 3,4,5,6 in \cref{fig:surface code stabilizer measurement four time steps}). Let $Z_{1}(k),\cdots, Z_{4}(k)$ ($X_{1}(k),\cdots, X_{4}(k)$) be the label of the data GKP qubit that the $k^{\textrm{th}}$ $Z$-type ($X$-type) syndrome GKP qubit is coupled with in Steps $3,\cdots ,6$. (If the syndrome GKP qubit is idling, the value is set to be zero). For example when $d=3$, $Z_{1}(k)$ and $X_{1}(k)$ are given by 
\begin{alignat}{4}
&Z_{1}(1) = 1,\,\,  & Z_{1}(2) &= 3,\,\, & Z_{1}(3) &= 5,\,\, & Z_{1}(4) &= 0, 
\nonumber\\
&X_{1}(1) = 2,\,\,  & X_{1}(2) &= 0,\,\, & X_{1}(3) &= 8,\,\, & X_{1}(4) &= 6, 
\end{alignat}
representing the connectivity between the syndrome and the data GKP qubits in Step 3.  

Due to the shift errors occurring during the preparation of GKP states, the noise vectors are updated as follows: 
\begin{align}
\xi_{q}^{(Dk)} &\leftarrow \xi_{q}^{(Dk)}  + \textrm{randG}(\sigma^{2}), 
\nonumber\\
\xi_{p}^{(Dk)} &\leftarrow \xi_{p}^{(Dk)}  + \textrm{randG}(\sigma^{2}),
\nonumber\\
\xi_{q}^{(Z\ell )} &\leftarrow  \textrm{randG}(\sigma_{\textrm{gkp}}^{2} ), 
\nonumber\\
\xi_{p}^{(Z\ell )} &\leftarrow  \textrm{randG}(\sigma_{\textrm{gkp}}^{2} ),
\nonumber\\
\xi_{q}^{(X\ell )} &\leftarrow  \textrm{randG}(\sigma_{\textrm{gkp}}^{2} ), 
\nonumber\\
\xi_{p}^{(X\ell )} &\leftarrow  \textrm{randG}(\sigma_{\textrm{gkp}}^{2} ), \label{eq:noise during gkp preparation surface code} 
\end{align}
for $k \in\lbrace 1,\cdots, d^{2} \rbrace$ and $\ell \in \lbrace 1,\cdots, d'  \rbrace$. In Step 3, the SUM gates transform the noise vectors as 
\begin{align}
(\xi_{q}^{(DZ_{1}(\ell))} , \xi_{q}^{(Z\ell)} ) &\leftarrow  (\xi_{q}^{(DZ_{1}(\ell))} , \xi_{q}^{(Z\ell)} +\xi_{q}^{(DZ_{1}(\ell))} ) 
\nonumber\\
&\qquad + \textrm{randG}\Big{(} \sigma^{2}\begin{bmatrix}
1 & 1/2 \\
1/2 & 4/3
\end{bmatrix} \Big{)} , 
\nonumber\\
(\xi_{p}^{(DZ_{1}(\ell))} , \xi_{p}^{(Z\ell)} ) &\leftarrow  (\xi_{p}^{(DZ_{1}(\ell))} - \xi_{p}^{(Z\ell)} , \xi_{p}^{(Z\ell)} ) 
\nonumber\\
&\qquad + \textrm{randG}\Big{(} \sigma^{2}\begin{bmatrix}
4/3 & -1/2 \\
-1/2 & 1
\end{bmatrix} \Big{)} , \label{eq:SUM gates between Z-type and data step 3}
\end{align}
for all $\ell \in \lbrace 1,\cdots, d' \rbrace$ if $Z_{1}(\ell)\neq 0$ and 
\begin{align}
\xi_{q}^{(Z\ell)} &\leftarrow \xi_{q}^{(Z\ell)} + \textrm{randG}(\sigma^{2}),
\nonumber\\
\xi_{p}^{(Z\ell)} &\leftarrow \xi_{p}^{(Z\ell)} + \textrm{randG}(\sigma^{2}), \label{eq:SUM gates between Z-type and data step 3 idling Z-type}
\end{align}
if $Z_{1}(\ell) =  0$. Similarly, 
\begin{align}
(\xi_{q}^{(DX_{1}(\ell))} , \xi_{q}^{(X\ell)} ) &\leftarrow  (\xi_{q}^{(DX_{1}(\ell))} + \xi_{q}^{(X\ell)}  , \xi_{q}^{(X\ell)}  ) 
\nonumber\\
&\qquad + \textrm{randG}\Big{(} \sigma^{2}\begin{bmatrix}
4/3 & 1/2 \\
1/2 & 1
\end{bmatrix} \Big{)} , 
\nonumber\\
(\xi_{p}^{(DX_{1}(\ell))} , \xi_{p}^{(X\ell)} ) &\leftarrow  (\xi_{p}^{(DX_{1}(\ell))}  , \xi_{p}^{(X\ell)} -\xi_{p}^{(DX_{1}(\ell))} ) 
\nonumber\\
&\quad + \textrm{randG}\Big{(} \sigma^{2}\begin{bmatrix}
1 & -1/2 \\
-1/2 & 4/3
\end{bmatrix} \Big{)} , \label{eq:SUM gates between $X$-type and data step 3}
\end{align}
for all $\ell \in \lbrace 1,\cdots, d' \rbrace$ if $X_{1}(\ell)\neq 0$ and 
\begin{align}
\xi_{q}^{(X\ell)} &\leftarrow \xi_{q}^{(X\ell)} + \textrm{randG}(\sigma^{2}),
\nonumber\\
\xi_{p}^{(X\ell)} &\leftarrow \xi_{p}^{(X\ell)} + \textrm{randG}(\sigma^{2}), \label{eq:SUM gates between $X$-type and data step 3 idling $X$-type}
\end{align}
if $X_{1}(\ell) = 0$. Since there are idling data GKP qubits, the data noise vectors are updated as 
\begin{align}
\xi_{q}^{(Dk)} &\leftarrow \xi_{q}^{(Dk)} + \textrm{randG}(\sigma^{2}), 
\nonumber\\
\xi_{p}^{(Dk)} &\leftarrow \xi_{p}^{(Dk)} + \textrm{randG}(\sigma^{2}), \label{eq:idling data step 3}
\end{align}
only for $k$ such that $Z_{1}(\ell)\neq k$ and $X_{1}(\ell)\neq k$ for all $\ell \in \lbrace 1,\cdots ,d' \rbrace$.

In Step 4, the SUM gates between the $Z$-type syndrome GKP qubits and data GKP qubits transform the noise vectors in the same way as in \cref{eq:SUM gates between Z-type and data step 3,eq:SUM gates between Z-type and data step 3 idling Z-type} except that $Z_{1}(\ell)$ is replaced by $Z_{2}(\ell)$. However, since the $X$-type syndrome GKP qubits are coupled with the data GKP qubits through inverse-SUM gates instead of SUM gates, the noise vectors are then updated as 
\begin{align}
(\xi_{q}^{(DX_{2}(\ell))} , \xi_{q}^{(X\ell)} ) &\leftarrow  (\xi_{q}^{(DX_{2}(\ell))} - \xi_{q}^{(X\ell)}  , \xi_{q}^{(X\ell)}  ) 
\nonumber\\
&\qquad + \textrm{randG}\Big{(} \sigma^{2}\begin{bmatrix}
4/3 & -1/2 \\
-1/2 & 1
\end{bmatrix} \Big{)} , 
\nonumber\\
(\xi_{p}^{(DX_{2}(\ell))} , \xi_{p}^{(X\ell)} ) &\leftarrow  (\xi_{p}^{(DX_{2}(\ell))}  , \xi_{p}^{(X\ell)} +\xi_{p}^{(DX_{2}(\ell))} ) 
\nonumber\\
&\quad + \textrm{randG}\Big{(} \sigma^{2}\begin{bmatrix}
1 & 1/2 \\
1/2 & 4/3
\end{bmatrix} \Big{)} , \label{eq:inverse-SUM gates between $X$-type and data step 4}
\end{align}
for all $\ell \in \lbrace 1,\cdots, d' \rbrace$ if $X_{2}(\ell)\neq 0$ and 
\begin{align}
\xi_{q}^{(X\ell)} &\leftarrow \xi_{q}^{(X\ell)} + \textrm{randG}(\sigma^{2}),
\nonumber\\
\xi_{p}^{(X\ell)} &\leftarrow \xi_{p}^{(X\ell)} + \textrm{randG}(\sigma^{2}), \label{eq:inverse-SUM gates between $X$-type and data step 4 idling Z-type}
\end{align}
if $X_{2}(\ell) = 0$, instead of as in \cref{eq:SUM gates between $X$-type and data step 3,eq:SUM gates between $X$-type and data step 3 idling $X$-type}. Due to the idling data GKP qubits, the noise vectors are further updated as in \cref{eq:idling data step 3} only for $k$ such that $Z_{2}(\ell)\neq k$ and $X_{2}(\ell)\neq k$ for all $\ell \in \lbrace 1,\cdots ,d' \rbrace$.

\begin{figure*}
\centering
\includegraphics[width=6.4in]{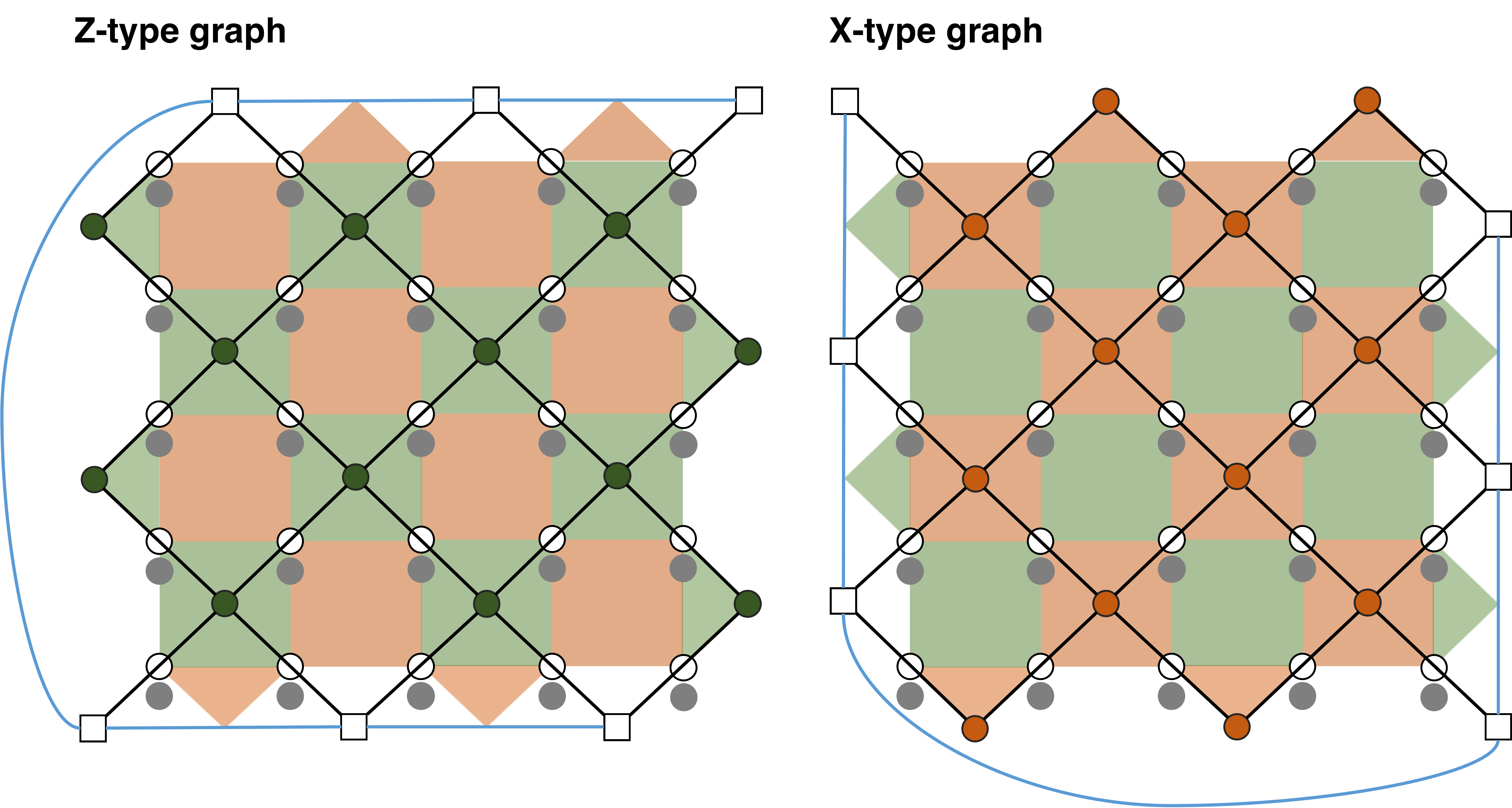}
\caption{$Z$-type and $X$-type 2D space graphs for the surface-GKP code with $d=5$. These 2D graphs will be stacked up to construct $Z$-type and $X$-type 3D space-time graphs. }
\label{fig:MWPM graphs} 
\end{figure*}

Note that in Step 5 and Step 6, the $X$-type syndrome GKP qubits are coupled with the data GKP qubits via inverse-SUM gates and SUM gates, respectively. Therefore, in Step 5, the noise vectors are updated in the same way as in Step 4, except that $Z_{2}(\ell)$ and $Z_{2}(\ell)$ are replaced by $Z_{3}(\ell)$ and $X_{3}(\ell)$. On the other hand, in Step 6, the noise vectors are updated in the same way as in Step 3, except that $Z_{1}(\ell)$ and $X_{1}(\ell)$ are replaced by $Z_{4}(\ell)$ and $X_{4}(\ell)$. Due to the noise before (or during) the homodyne measurement, the noise vectors are updated as 
\begin{align}
\xi_{q}^{(Dk)} &\leftarrow \xi_{q}^{(Dk)}  + \textrm{randG}(\sigma^{2}), 
\nonumber\\
\xi_{p}^{(Dk)} &\leftarrow \xi_{p}^{(Dk)}  + \textrm{randG}(\sigma^{2}),
\nonumber\\
\xi_{q}^{(Z\ell )} &\leftarrow \xi_{q}^{(Z\ell )} +  \textrm{randG}(\sigma^{2} ), 
\nonumber\\
\xi_{p}^{(Z\ell )} &\leftarrow \xi_{p}^{(Z\ell )} +  \textrm{randG}(\sigma^{2} ), 
\nonumber\\
\xi_{q}^{(X\ell )} &\leftarrow \xi_{q}^{(X\ell )} +  \textrm{randG}(\sigma^{2} ), 
\nonumber\\
\xi_{p}^{(X\ell )} &\leftarrow \xi_{p}^{(X\ell )} +  \textrm{randG}(\sigma^{2} ),  \label{eq:noise during homodyne measurement surface code}
\end{align}
for all $k\in\lbrace 1,\cdots, d^{2} \rbrace$ and $\ell\in\lbrace 1,\cdots, d' \rbrace$. Then, through the homodyne measurement, we measure $\xi_{q}^{(Z\ell )}$ and $\xi_{p}^{(X\ell )}$ modulo $2\sqrt{\pi}$ and assign stabilizer values as 
\begin{align}
\hat{S}_{Z}^{(\ell)} &\leftarrow \begin{cases}
+1 & | R_{\sqrt{2\pi}}( \xi_{q}^{(Z\ell )} )  |  \le \sqrt{\pi}/2  \\
-1 & | R_{\sqrt{2\pi}}( \xi_{q}^{(Z\ell )} )  |  > \sqrt{\pi}/2 
\end{cases}, 
\nonumber\\
\hat{S}_{X}^{(\ell)} &\leftarrow \begin{cases}
+1 & | R_{\sqrt{2\pi}}( \xi_{p}^{(X\ell )} )  |  \le \sqrt{\pi}/2  \\
-1 & | R_{\sqrt{2\pi}}( \xi_{p}^{(X\ell )} )  |  > \sqrt{\pi}/2 
\end{cases}, \label{eq:stabilizer value assignment}
\end{align}
for all $\ell \in \lbrace 1,\cdots, d' \rbrace$. $R_{s}(z)$ is defined in \cref{eq:definition of the R function}.  

\subsection*{Construction of 3D space-time graphs} 
\label{subsection:Construction of the 3D space-time graphs}

Now we construct 3D space-time graphs to which we will apply a minimum-weight perfect matching decoding algorithm. The overall structure is as follows: Since each stabilizer measurement can be faulty, we repeat the noisy stabilizer measurement cycle $d$ times. Then, we perform another round of ideal stabilizer measurement cycle assuming that all circuit elements and supplied GKP states are noiseless. The reason for adding the extra noiseless measurement cycle is to ensure that the noisy states are restored back to the code space so we can later conveniently determine whether the error correction succeed or not. Then, the $Z$-type and the $X$-type 3D space-time graphs are constructed to represent the outcomes of $d+1$ rounds of stabilizer measurement cycles. These space-time graphs will then be used to decode the $Z$-type and the $X$-type syndrome measurement outcomes. 

We first construct the $Z$-type and $X$-type 2D space graphs as in \cref{fig:MWPM graphs}. Each bulk vertex of the 2D space graph corresponds to a syndrome GKP qubit and each bulk edge corresponds to a data GKP qubit. Note also that there are boundary vertices (squares in \cref{fig:MWPM graphs}) that do not correspond to any syndrome GKP qubits and the corresponding boundary edges (blue lines in \cref{fig:MWPM graphs}) that are not associated with any data GKP qubits. Therefore, the boundary edge weighs are always set to be zero. 

Then, we associate each 2D space graph with one round of stabilizer measurement cycle. So, there are $d+1$ 2D space graphs and these 2D space graphs are stacked up together by introducing vertical edges that connect the same vertices in two adjacent 2D space graphs (corresponding to two adjacent stabilizer measurement rounds). Below, we discuss in detail how the bulk edge weights are assigned.

We start by initializing the data position and momentum noise vectors to a zero vector: 
\begin{align}
\xi_{q}^{D} &= (\xi_{q}^{(D1)}, \cdots, \xi_{q}^{(Dd^{2})}) = (0,\cdots, 0), 
\nonumber\\
\xi_{p}^{D} &= (\xi_{p}^{(D1)}, \cdots, \xi_{p}^{(Dd^{2})}) = (0,\cdots, 0). 
\end{align} 
These data noise vectors are fed into Step 1 of GKP-stabilizer measurement as described in \cref{eq:noise during gkp preparation,eq:SUM between data and ancilla,eq:inverse-SUM between data and ancilla,eq:noise during homodyne measurement}. Let $w_{Z}^{H}(k)$ and $w_{Z}^{H}(k)$ be the horizontal edge weights of the $Z$-type and $X$-type graphs corresponding to the $k^{\textrm{th}}$ data GKP qubit ($k\in\lbrace 1,\cdots, d^{2} \rbrace$). Then, while updating the data position and momentum noise vectors as prescribed in \cref{eq:position correction,eq:momentum correction}, we assign the horizontal edge weights as
\begin{widetext}
\begin{align}
w_{Z}^{H}(k) &\leftarrow \begin{cases}
-\log_{2}\big{(} p[ \sqrt{\sigma_{\textrm{gkp}}^{2} + \frac{10}{3}\sigma^{2}} ]\big{(} R_{\sqrt{\pi}}( \xi_{q}^{(Ak)} ) \big{)} \big{)} & \textrm{round }1 \\
-\log_{2}\big{(} p[ \sigma_{Z}^{H}(k;d) ]\big{(} R_{\sqrt{\pi}}( \xi_{q}^{(Ak)} ) \big{)} \big{)} &  \textrm{round }2\textrm{ to round }d\\
-\log_{2}\big{(} p[ \sqrt{ (\sigma_{Z}^{H}(k;d) )^{2} - \sigma_{\textrm{gkp}}^{2} - \frac{10}{3}\sigma^{2}  } ]\big{(} R_{\sqrt{\pi}}( \xi_{q}^{(Ak)} )   \big{)} \big{)} & \textrm{round }d+1
\end{cases}, 
\end{align}
for odd $k$ and 
\begin{align}
w_{X}^{H}(k) &\leftarrow \begin{cases}
-\log_{2}\big{(} p[ \sqrt{\sigma_{\textrm{gkp}}^{2} + \frac{10}{3}\sigma^{2}} ]\big{(} R_{\sqrt{\pi}}( \xi_{p}^{(Ak)} ) \big{)} \big{)} & \textrm{round }1 \\
-\log_{2}\big{(} p[ \sigma_{X}^{H}(k;d) ]\big{(} R_{\sqrt{\pi}}( \xi_{p}^{(Ak)} ) \big{)} \big{)} &  \textrm{round }2\textrm{ to round }d\\
-\log_{2}\big{(} p[ \sqrt{ (\sigma_{X}^{H}(k;d) )^{2} - \sigma_{\textrm{gkp}}^{2} - \frac{10}{3}\sigma^{2}  } ]\big{(} R_{\sqrt{\pi}}( \xi_{p}^{(Ak)} )   \big{)} \big{)} & \textrm{round }d+1
\end{cases}, 
\end{align}
for even $k$ if the additional GKP information is used. Here, we use $\xi_{q}^{(Ak)}$ and $\xi_{p}^{(Ak)}$ that are obtained after applying \cref{eq:noise during homodyne measurement}. $p_{\textrm{err}}(\sigma)$ and $p[\sigma](z)$ are defined in \cref{eq:definition of perr,eq:Definition of the p function} and $R_{s}(z)$ is defined in \cref{eq:definition of the R function}. On the other hand, if the additional GKP information is not used, we assign the horizontal edge weights as 
\begin{align}
w_{Z}^{H}(k) &\leftarrow \begin{cases}
-\log_{2}\big{(} p_{\textrm{err}} \big{(}  \sqrt{\sigma_{\textrm{gkp}}^{2} + \frac{10}{3}\sigma^{2}} \big{)} \big{)} & \textrm{round }1 \\
-\log_{2}\big{(} p_{\textrm{err}} \big{(} \sigma_{Z}^{H}(k;d) \big{)} \big{)} &  \textrm{round }2\textrm{ to round }d\\
-\log_{2}\big{(} p_{\textrm{err}} \big{(} \sqrt{ (\sigma_{Z}^{H}(k;d) )^{2} - \sigma_{\textrm{gkp}}^{2} - \frac{10}{3}\sigma^{2}  } \big{)} \big{)} & \textrm{round }d+1
\end{cases}, 
\end{align}
for odd $k$ and 
\begin{align}
w_{X}^{H}(k) &\leftarrow \begin{cases}
-\log_{2}\big{(} p_{\textrm{err}} \big{(} \sqrt{\sigma_{\textrm{gkp}}^{2} + \frac{10}{3}\sigma^{2}} \big{)} \big{)}  & \textrm{round }1 \\
-\log_{2}\big{(} p_{\textrm{err}} \big{(} \sigma_{X}^{H}(k;d) \big{)} \big{)}  &  \textrm{round }2\textrm{ to round }d\\
-\log_{2}\big{(} p_{\textrm{err}} \big{(} \sqrt{ (\sigma_{X}^{H}(k;d) )^{2} - \sigma_{\textrm{gkp}}^{2} - \frac{10}{3}\sigma^{2}  } \big{)} \big{)} & \textrm{round }d+1
\end{cases}, 
\end{align}
for even $k$. Here, $\sigma_{Z}^{H}(k;d)$ and $\sigma_{X}^{H}(k;d)$ are defined as 
\begin{align}
\sigma_{Z}^{H}(k;d) &\equiv \begin{cases}
\begin{cases}
\sqrt{  4\sigma_{\textrm{gkp}}^{2} + \frac{52}{3}\sigma^{2} }  & \frac{k-1}{d}\in 2\mathbb{Z} \\
\sqrt{  4\sigma_{\textrm{gkp}}^{2} + \frac{58}{3}\sigma^{2} }  & \frac{k-1}{d}\in 2\mathbb{Z}+1
\end{cases} & k\in d\mathbb{Z} +1 \\
\begin{cases}
\sqrt{  4\sigma_{\textrm{gkp}}^{2} + \frac{55}{3}\sigma^{2} }  & \frac{k}{d}\in 2\mathbb{Z}+1 \\
\sqrt{  4\sigma_{\textrm{gkp}}^{2} + \frac{49}{3}\sigma^{2} }  & \frac{k}{d}\in 2\mathbb{Z}
\end{cases} & k\in d\mathbb{Z} \\
\sqrt{  5\sigma_{\textrm{gkp}}^{2} + \frac{59}{3}\sigma^{2} } & \textrm{otherwise}
\end{cases}, 
\nonumber\\
\sigma_{X}^{H}(k;d) &\equiv \begin{cases}
\begin{cases}
\sqrt{  4\sigma_{\textrm{gkp}}^{2} + \frac{49}{3}\sigma^{2} } & k\in 2\mathbb{Z} + 1  \\
\sqrt{  4\sigma_{\textrm{gkp}}^{2} + \frac{55}{3}\sigma^{2} } & k\in 2\mathbb{Z} 
\end{cases} &  k\in \lbrace 1,\cdots, d \rbrace \\
\begin{cases}
\sqrt{  4\sigma_{\textrm{gkp}}^{2} + \frac{58}{3}\sigma^{2} } & k\in 2\mathbb{Z} + 1  \\
\sqrt{  4\sigma_{\textrm{gkp}}^{2} + \frac{52}{3}\sigma^{2} } & k\in 2\mathbb{Z} 
\end{cases} &  k\in \lbrace d^{2}-d+1 ,\cdots, d^{2} \rbrace \\
\sqrt{  5\sigma_{\textrm{gkp}}^{2} + \frac{59}{3}\sigma^{2} }&\textrm{otherwise}
\end{cases}. 
\end{align}
We remark that we have carefully determined $\sigma_{Z}^{H}(k;d)$ and $\sigma_{X}^{H}(k;d)$ by thoroughly keeping tracking of how the circuit-level noise propagates   

Then, moving on to Step 2 of GKP-stabilizer measurement, we update the noise vectors as described in \cref{eq:noise during gkp preparation,eq:SUM between data and ancilla,eq:inverse-SUM between data and ancilla,eq:noise during homodyne measurement}, except that \cref{eq:SUM between data and ancilla,eq:inverse-SUM between data and ancilla} are applied for even and odd $k$ (instead of odd and even $k$), respectively. Similarly as above, while updating the data position and momentum noise vectors as prescribed in \cref{eq:position correction,eq:momentum correction}, we assign the horizontal edge weights as 
\begin{align}
w_{Z}^{H}(k) &\leftarrow \begin{cases}
-\log_{2}\big{(} p[ \sqrt{\sigma_{2\textrm{gkp}}^{2} + \frac{20}{3}\sigma^{2}} ]\big{(} R_{\sqrt{\pi}}( \xi_{q}^{(Ak)} ) \big{)} \big{)} & \textrm{round }1 \\
-\log_{2}\big{(} p[ \sigma_{Z}^{H}(k;d) ]\big{(} R_{\sqrt{\pi}}( \xi_{q}^{(Ak)} ) \big{)} \big{)} &  \textrm{round }2\textrm{ to round }d\\
-\log_{2}\big{(} p[ \sqrt{ (\sigma_{Z}^{H}(k;d) )^{2} - 2\sigma_{\textrm{gkp}}^{2} - \frac{20}{3}\sigma^{2}  } ]\big{(} R_{\sqrt{\pi}}( \xi_{q}^{(Ak)} )   \big{)} \big{)} & \textrm{round }d+1
\end{cases}, 
\end{align}
for even $k$ and 
\begin{align}
w_{X}^{H}(k) &\leftarrow \begin{cases}
-\log_{2}\big{(} p[ \sqrt{2\sigma_{\textrm{gkp}}^{2} + \frac{20}{3}\sigma^{2}} ]\big{(} R_{\sqrt{\pi}}( \xi_{p}^{(Ak)} ) \big{)} \big{)} & \textrm{round }1 \\
-\log_{2}\big{(} p[ \sigma_{X}^{H}(k;d) ]\big{(} R_{\sqrt{\pi}}( \xi_{p}^{(Ak)} ) \big{)} \big{)} &  \textrm{round }2\textrm{ to round }d\\
-\log_{2}\big{(} p[ \sqrt{ (\sigma_{X}^{H}(k;d) )^{2} - 2\sigma_{\textrm{gkp}}^{2} - \frac{20}{3}\sigma^{2}  } ]\big{(} R_{\sqrt{\pi}}( \xi_{p}^{(Ak)} )   \big{)} \big{)} & \textrm{round }d+1
\end{cases}, 
\end{align}
for odd $k$ if the additional GKP information is used. Here, we use $\xi_{q}^{(Ak)}$ and $\xi_{p}^{(Ak)}$ that are obtained after applying \cref{eq:noise during homodyne measurement}. If on the other hand the additional GKP information is not used, we assign the horizontal edge weights as 
\begin{align}
w_{Z}^{H}(k) &\leftarrow \begin{cases}
-\log_{2}\big{(} p_{\textrm{err}} \big{(}  \sqrt{2\sigma_{\textrm{gkp}}^{2} + \frac{20}{3}\sigma^{2}} \big{)} \big{)} & \textrm{round }1 \\
-\log_{2}\big{(} p_{\textrm{err}} \big{(} \sigma_{Z}^{H}(k;d) \big{)} \big{)} &  \textrm{round }2\textrm{ to round }d\\
-\log_{2}\big{(} p_{\textrm{err}} \big{(} \sqrt{ (\sigma_{Z}^{H}(k;d) )^{2} - 2\sigma_{\textrm{gkp}}^{2} - \frac{20}{3}\sigma^{2}  } \big{)} \big{)} & \textrm{round }d+1
\end{cases}, 
\end{align}
for even $k$ and 
\begin{align}
w_{X}^{H}(k) &\leftarrow \begin{cases}
-\log_{2}\big{(} p_{\textrm{err}} \big{(} \sqrt{2\sigma_{\textrm{gkp}}^{2} + \frac{20}{3}\sigma^{2}} \big{)} \big{)}  & \textrm{round }1 \\
-\log_{2}\big{(} p_{\textrm{err}} \big{(} \sigma_{X}^{H}(k;d) \big{)} \big{)} &  \textrm{round }2\textrm{ to round }d\\
-\log_{2}\big{(} p_{\textrm{err}} \big{(} \sqrt{ (\sigma_{X}^{H}(k;d) )^{2} - 2\sigma_{\textrm{gkp}}^{2} - \frac{20}{3}\sigma^{2}  } \big{)} \big{)} & \textrm{round }d+1
\end{cases}, 
\end{align}
for odd $k$. This way, all the horizontal edge weights are assigned. 

Vertical edge weights are assigned during surface code stabilizer measurements: We follow Steps 3--6 of surface code stabilizer measurements and update the noise vectors as described in \cref{eq:noise during gkp preparation surface code} to \cref{eq:noise during homodyne measurement surface code}. Let $w_{Z}^{V}(\ell)$ and $w_{X}^{V}(\ell)$ be the vertical edge weights of the $Z$-type and $X$-type 3D space-time graphs corresponding to the $\ell^{\textrm{th}}$ $Z$-type and $X$-type syndrome qubit. Then, after assigning the stabilizer values as in \cref{eq:stabilizer value assignment}, we further assign the vertical edge weights as follows:  
\begin{align}
w_{Z}^{V}(\ell) &\leftarrow  -\log_{2}\big{(} p[\sigma_{Z}^{V}(\ell;d)] \big{(} R_{\sqrt{\pi}}( \xi_{q}^{(Zk)} ) \big{)} \big{)}, 
\nonumber\\
w_{X}^{V}(\ell) &\leftarrow  -\log_{2}\big{(} p[\sigma_{X}^{V}(\ell;d)] \big{(} R_{\sqrt{\pi}}( \xi_{p}^{(Xk)} ) \big{)} \big{)}, 
\end{align}
while in rounds $1$ to $d$ for all $\ell \in\lbrace 1,\cdots, d' = (d^{2}-1)/2 \rbrace$, if the additional GKP information is used. Here, we use $\xi_{q}^{(Zk)}$ and $\xi_{p}^{(Xk)}$ that are obtained after applying \cref{eq:noise during homodyne measurement surface code} and $\sigma_{Z}^{V}(\ell;d)$ and $\sigma_{X}^{V}(\ell;d)$ are defined as 
\begin{align}
\sigma_{Z}^{V}(\ell;d)&= \begin{cases}
\sqrt{ 4\sigma_{\textrm{gkp}}^{2} + \frac{56}{3}\sigma^{2}  }  & \ell\in 2d'' \mathbb{Z} + 1 \\
\sqrt{ 7\sigma_{\textrm{gkp}}^{2} + \frac{107}{3}\sigma^{2}  }  & \ell\in 2d'' \mathbb{Z} + d''+ 1 \\
\sqrt{ 4\sigma_{\textrm{gkp}}^{2} + \frac{73}{3}\sigma^{2}  }  & \ell\in 2d'' \mathbb{Z} \\
\sqrt{ 7\sigma_{\textrm{gkp}}^{2} + \frac{116}{3}\sigma^{2}  }&\textrm{otherwise}
\end{cases}, 
\nonumber\\
\sigma_{X}^{V}(\ell;d)&= \begin{cases}
\sqrt{  4\sigma_{\textrm{gkp}}^{2} + \frac{56}{3}\sigma^{2}  } & \ell\in 2d''\mathbb{Z} + d''  \\
\sqrt{  4\sigma_{\textrm{gkp}}^{2} + \frac{73}{3}\sigma^{2}  } & \ell\in 2d''\mathbb{Z} +d''+ 1 \\
\sqrt{  7\sigma_{\textrm{gkp}}^{2} + \frac{107}{3}\sigma^{2}  } & \ell\in 2d''\mathbb{Z}  \\
\sqrt{  7\sigma_{\textrm{gkp}}^{2} + \frac{116}{3}\sigma^{2}  } & \textrm{otherwise}\\
\end{cases}. 
\end{align}
Similarly as above, we have carefully determined $\sigma_{Z}^{V}(\ell;d)$ and $\sigma_{X}^{V}(\ell;d)$ by thoroughly keeping track of how the circuit-level noise propagates. If on the other hand the additional GKP information is not used, we assign the vertical edge weights as 
\begin{align}
w_{Z}^{V}(\ell) &\leftarrow  -\log_{2}\big{(} p_{\textrm{err}} \big{(} \sigma_{Z}^{V}(\ell;d) \big{)} \big{)}, 
\nonumber\\
w_{X}^{V}(\ell) &\leftarrow  -\log_{2}\big{(} p_{\textrm{err}} \big{(} \sigma_{X}^{V}(\ell;d) \big{)}  \big{)}. 
\end{align}
This way, all the vertical edge weights are assigned and thus we are left with the complete $Z$-type and $X$-type 3D space-time graphs with all the horizontal and vertical edge weights assigned.  
\end{widetext}

\subsection*{Minimum-weight perfect matching decoding}

Now, given the 3D space-time graphs, the correction is determined by using a minimum-weight perfect matching decoding algorithm. More specifically, we do the following: 
\begin{enumerate}
\item Simulate $d$ rounds of noisy stabilizer measurements followed by one round of ideal stabilizer measurements and construct the $Z$-type and $X$-type 3D space-time graphs as described above. 
\item Highlight all vertices whose assigned stabilizer value is changed from the previous round. If the number of highlighted vertices is odd, highlight a boundary vertex. Thus, the number of highlighted vertices is always even. 
\item For all pairs of highlighted $Z$-type ($X$-type) vertices, find the path with the minimum total weight. Then, save the minimum total weight and all edges in the path. Then, we are left with a $Z$-type ($X$-type) complete graph of highlighted vertices, where the weight of the edge $(v,w)$ is given by the minimum total weight of the path that connects $v$ and $w$. 
\item Apply the minimum-weight perfect matching algorithm \cite{Edmonds1965,Edmonds1965b} on the $Z$-type ($X$-type) complete graph of highlighted vertices. For all matched pairs of $Z$-type ($X$-type) vertices, highlight all the $Z$-type ($X$-type) edges contained in the path that connects the matched vertices. 
\item Suppress all vertical edges and project the $Z$-type ($X$-type) 3D space-time graph onto the 2D plane. For each $Z$-type ($X$-type) horizontal edge, count how many times it was highlighted. If it is highlighted even times, do nothing. Otherwise, apply the Pauli correction operator $\hat{X}_{\textrm{gkp}}$ ($\hat{Z}_{\textrm{gkp}}$) to the corresponding data GKP qubit. Equivalently, update the quadrature noise as $\xi_{q}^{(Dk)}\leftarrow \xi_{q}^{(Dk)} +\sqrt{\pi}$ ($\xi_{p}^{(Dk)}\leftarrow \xi_{p}^{(Dk)} +\sqrt{\pi}$). 
\end{enumerate}

Once the correction is done, we are left with the data noise vectors $\xi_{q}^{D} = (\xi_{q}^{(D1)},\cdots ,\xi_{q}^{(Dd^{2})})$ and $\xi_{p}^{D}=(\xi_{p}^{(D1)},\cdots ,\xi_{p}^{(Dd^{2})})$. Define $\textrm{total}(\xi_{q}^{D}) \equiv \sum_{k=1}^{d^{2}}\xi_{q}^{(Dk)} $ and $\textrm{total}(\xi_{p}^{D}) \equiv \sum_{k=1}^{d^{2}}\xi_{p}^{(Dk)} $. Then, we determine that there is 
\begin{align}
\begin{cases}
\textrm{logical }X & \textrm{total}(\xi_{q}^{D}) = \textrm{odd }\& \textrm{ total}(\xi_{p}^{D}) =\textrm{even}\\
\textrm{logical }Z & \textrm{total}(\xi_{q}^{D})= \textrm{even }\& \textrm{ total}(\xi_{p}^{D}) =\textrm{odd}\\
\textrm{logical }Y & \textrm{total}(\xi_{q}^{D}) = \textrm{odd }\& \textrm{ total}(\xi_{p}^{D}) =\textrm{odd}
\end{cases}
\end{align} 
error. Otherwise if both $\textrm{total}(\xi_{q}^{D})$ and $\textrm{total}(\xi_{p}^{D})$ are even, there is no logical error. 

We use the Monte Carlo method to compute the logical $X,Y,Z$ error probability. In \cref{fig:main results}, we plot the logical $X$ error probability obtained from 10,000--100,000 samples, which is the same as the logical $Z$ error probability. The number of samples is determined such that statistical fluctuations are negligible.

\bibliography{GKP_surface_v10}






\end{document}